\documentclass[structabstract]{aa}  
\usepackage{graphicx}
\usepackage{dcolumn}
\usepackage{txfonts}
\usepackage{natbib}
\usepackage{natbib,twoopt}
\bibpunct{(}{)}{;}{a}{}{,} 
\newcommandtwoopt{\citeads}[3][][]{\href{http://adsabs.harvard.edu/abs/#3}%
{\citealp[#1][#2]{#3}}}
\newcommandtwoopt{\citepads}[3][][]{\href{http://adsabs.harvard.edu/abs/#3}%
{\citep[#1][#2]{#3}}}
\newcommandtwoopt{\citetads}[3][][]{\href{http://adsabs.harvard.edu/abs/#3}%
{\citet[#1][#2]{#3}}} 
\newcommandtwoopt{\citeyearads}[3][][]%
{\href{http://adsabs.harvard.edu/abs/#3}{\citeyear[#1][#2]{#3}}}

\begin{document}
\authorrunning{S. Mathur et al.}
\titlerunning{Magnetic activity of F stars observed by {\emph Kepler}}
   \title{Magnetic activity of F stars observed by {\it Kepler}}


  \author{S. Mathur\inst{1,2}
          \and
          R.~A. Garc\'ia\inst{1}
          \and
          J. Ballot\inst{3,4}
           \and
          T. Ceillier\inst{1}
          \and    
          D. Salabert\inst{1, 5}
          \and
         T.~S. Metcalfe\inst{2, 6}
          \and 
         C. R\'egulo\inst{7, 8}         
          \and
          A. Jim\'enez\inst{7, 8}
          \and
         S. Bloemen\inst{9, 10}     
          }
\institute{Laboratoire AIM, CEA/DSM-CNRS-Universit\'e Paris Diderot; IRFU/SAp, Centre de Saclay, 91191 Gif-sur-Yvette Cedex, France\\ \email{savita.mathur@gmail.com; rgarcia@cea.fr}
\and
Space Science Institute, 4750 Walnut street Suite\#205, Boulder, CO 80301, USA\\ \email{smathur@spacescience.org}
\and
CNRS, Institut de Recherche en Astrophysique et Plan\'etologie, 14 avenue Edouard Belin, 31400 Toulouse, France\\ 
\email{ jerome.ballot@irap.omp.eu} 
\and
Universit\'e de Toulouse, UPS-OMP, IRAP, 31400 Toulouse, France 
\and
Laboratoire Lagrange, UMR7293, Universit\'e de Nice Sophia-Antipolis, CNRS, Observatoire de la C\^ote d'Azur, BP 4229, 06304 Nice Cedex 4, France\\ \email{salabert@oca.eu}
\and
Stellar Astrophysics Centre, Aarhus University, DK-8000 Aarhus C, Denmark
\and
Instituto de Astrof\'isica de Canarias, E-38200 La Laguna, Tenerife, Spain\\
\email{crr@iac.es}
\and
Departamento de Astrof\'isica,  Universidad de La Laguna, E-38206 La Laguna, Tenerife, Spain
\and
Instituut voor Sterrenkunde, University of Leuven, Celestijnenlaan 200 D, B-3001 Leuven, Belgium
\and
Department of Astrophysics, IMAPP, Radboud University Nijmegen, PO Box 9010, NL-6500 GL, Nijmegen, The Netherlands
}
  
  \date{Received August 18, 2011; accepted }

 
  \abstract
    {The study of stellar activity is important because it can provide new constraints for dynamo models when combined with surface rotation rates and the depth of the convection zone. We know that the dynamo mechanism, which is believed to be the main process that rules the magnetic cycle of solar-like stars, results from the interaction between (differential) rotation, convection, and magnetic field. The \emph{Kepler} mission has already been collecting data for a large number of stars during four years allowing us to investigate magnetic stellar cycles.}
 {We investigated the \emph{Kepler} light curves to look for magnetic activity or even hints of magnetic activity cycles. Based on the photometric data we also looked for new magnetic indexes to characterise the magnetic activity of the stars.}
{We selected a sample of 22 solar-like F stars that have a rotation period shorter than 12 days. We performed a time-frequency analysis using the Morlet wavelet yielding a magnetic proxy for our sample of stars. We computed the magnetic index $S_{\rm ph}$ as the standard deviation of the whole time series and the index $ \langle S_{\rm ph} \rangle$, which is the mean of standard deviations measured in subseries of length five times the rotation period of the star. We defined new indicators, such as the contrast between high and low activity, to take into account the fact  that complete magnetic cycles are not observed for all the stars. We also inferred the Rossby number of the stars and studied their stellar background.}
{This analysis shows different types of behaviour in the 22 F stars. Two stars show behaviour very similar to magnetic activity cycles. Five stars show long-lived spots or active regions suggesting the existence of active longitudes. Two stars in our sample seem to have a decreasing or increasing trend in the temporal variation of the magnetic proxies. Finally, the last group of stars shows magnetic activity (with the presence of spots) but no sign of cycle. }
   {}


   \maketitle
%

\section{Introduction}
The magnetic activity of the Sun observed as a star shows clear signatures 
of rotational modulation superimposed on longer-term variations due to the 
11-year sunspot cycle. Similar behaviour can also be seen in other stars by 
tracking their magnetic variations spectroscopically with observations of 
the Ca~{\sc ii} H (396.8\,nm) and K (393.4\,nm) lines, or through careful 
photometric monitoring to reveal the brightness changes associated with 
dark spots and bright faculae \citep[e.g. see][]{2007ApJS..171..260L}. The most 
comprehensive spectroscopic survey for these variations was conducted at 
the Mount Wilson observatory over more than 30 years \citep{1978ApJ...226..379W,
1995ApJ...438..269B}, leading to the first large sample of magnetic activity and 
rotation data to help validate models of stellar dynamos.

The Mount Wilson sample established that both the mean activity level and 
the stellar cycle period seemed to scale with the Rossby number, which 
tracks the stellar rotation period normalized by the convective timescale 
\citep[as defined by][]{1984ApJ...279..763N}. Essentially, the shortest activity cycles were observed 
in the most rapidly rotating stars, while more slowly rotating stars 
generally showed cycle periods comparable to the Sun or longer. Subsequent 
analysis revealed two distinct relationships between rotation period and 
cycle period, with an ``active'' sequence of stars rotating faster than 
the Sun showing a stellar cycle every few hundred rotations and an 
``inactive'' sequence of more slowly rotating stars showing a stellar 
cycle every $\sim$100 rotations \citep{1999ApJ...524..295S}. Some stars with 
intermediate rotation periods (10-25 days) exhibited stellar cycles on 
both sequences simultaneously. This led \cite{2007ApJ...657..486B} to suggest 
that the two sequences might represent distinct dynamos in different 
regions of the star, driven either in a shear layer at the base of the 
convection zone (a so-called tachocline) or in a near-surface shear layer 
like that seen in helioseismic inversions \citep{ThoToo1996}.

Evidence of the existence of this second class of short-period stellar 
cycles has been growing over the past few years. \cite{2010ApJ...718L..19F} used 
helioseismic data to identify a quasi-biennial ($\sim$2 year) variation in 
the Sun, with an amplitude that appeared to be modulated by the dominant 
11-year signal. They speculated that buoyant magnetic flux, created near 
the tachocline during the maximum of the 11-year sunspot cycle, would rise 
toward the surface and pump up the amplitude of this short-period cycle. 
However, other explanations, like the beating between different configurations of the dynamo, 
were also given to justify this second modulation \citep{2013ApJ...765..100S}. 
\cite{2010Sci...329.1032G} found asteroseismic signatures of a short stellar cycle in 
the F2V star HD~49933 using observations from the CoRoT satellite. The 
solar-like pattern of anti-correlated changes in the oscillation 
frequencies and amplitudes suggested a cycle period longer than 120\,days, and 
the size of the shifts showed a similar frequency dependence to that observed in 
the Sun \citep{1990Natur.345..779L,2011A&A...530A.127S}. \cite{2010ApJ...723L.213M} discovered a 
record-breaking 1.6-year activity cycle in the F8V star $\iota$~Hor 
(HD\,17051), which has subsequently been confirmed with X-ray observations 
of the stellar corona \citep{2013A&A...553L...6S}. Most recently, \cite{2013ApJ...763L..26M}
documented dual magnetic cycles with periods of 3 years and 13 years in 
the K2V star $\epsilon$~Eri (HD\,22049), falling squarely on the two 
activity sequences identified by \citeauthor{1999ApJ...524..295S}.

The {\it Kepler} mission represents an unprecedented opportunity to study 
the short-period magnetic cycles that have been observed in some rapidly 
rotating F stars. The high precision time-series photometry collected 
every 30 minutes over the past four years can be used to measure rotation 
periods from spot modulation and to monitor the longer term brightness 
changes associated with the stellar cycle. Furthermore, for targets that 
have also been observed in short-cadence \citep{2010ApJ...713L.160G}, the one-minute 
sampling allows a characterisation of the star through asteroseismology -- 
including quantities such as the depth of the surface convection zone \citep[e.g.][]{2012ApJ...749..152M,2012AN....333.1040M} and 
radial differential rotation-- although for more evolved stars \citep{2012Natur.481...55B,2012ApJ...756...19D}, which provide useful constraints for dynamo 
models. The asteroseismic data can also be used to monitor the p-mode 
frequencies over time, allowing a search for the same pattern of changes 
that have been seen for the Sun and HD~49933 in response to their magnetic 
cycles. Rapidly rotating F stars are the ideal targets because they are 
expected to show the shortest cycle periods, and the size of the frequency 
variations is predicted to be larger than in the Sun \citep{2007MNRAS.379L..16M}, 
as has also been suggested by observations \citep{2010Sci...329.1032G,2011A&A...530A.127S}.
 In addition, the F star tau Boo belongs to the handful of stars where a magnetic activity cycle
has been detected through Zeeman measurement of its magnetic field \citep{2009MNRAS.398.1383F}.

In this paper, we aim to characterise magnetic activity in a sample of 
rapidly rotating F stars using photometric measurements from the {\it 
Kepler} mission. Our goal is to better understand the development 
of regular magnetic cycles, and to place strong constraints on the 
physical processes that govern them. Our sample includes F stars with 
surface rotation periods shorter than 12 days that have also been observed 
with short-cadence for at least 3 months continuously, where the p-mode 
oscillations and rotation are both clearly detected. We provide an 
overview of the observations in Sect.~2 and define a new photometric 
index of magnetic activity in Sect.~3. We present our results in Sect.~
4 and discuss them in Sect.~5. A summary of our conclusions is presented 
in Sect.~6.

\section{Observations and Data Analysis}

\begin{figure*}[htb]
\begin{center}
\includegraphics[width=11cm, trim=2cm 6cm 4cm 0.5cm]{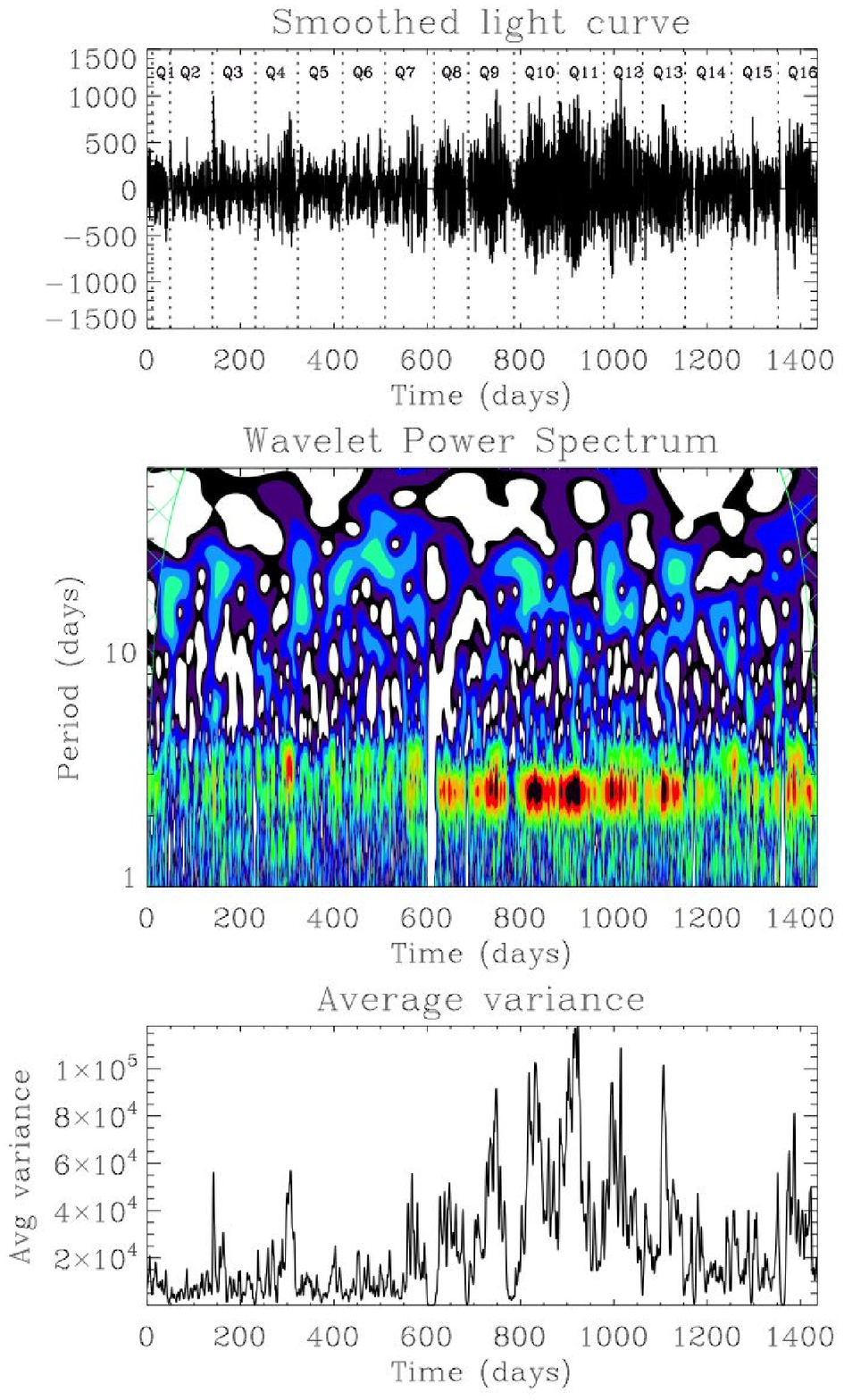}\\
\includegraphics[angle=90, width=7.2cm, trim=1cm 1cm -0.5cm 4cm]{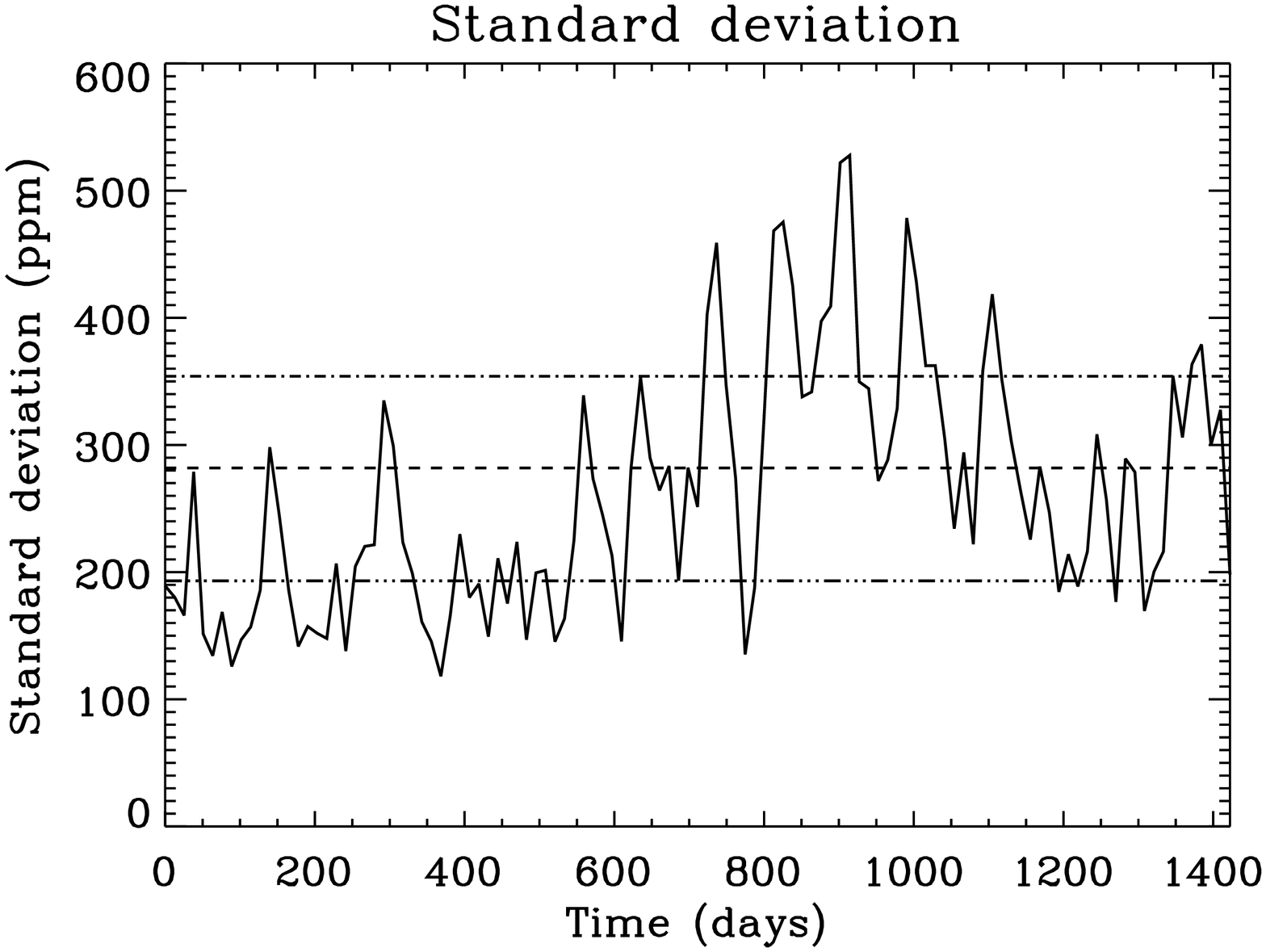}
\caption{ Wavelet analysis of KIC~3733735 (groups C and L). Top panel: temporal variation of the flux after the corrections applied as in \citet{2011MNRAS.414L...6G} and rebinned to 2 hrs. Middle panel: Wavelet power spectrum as a function of time and period. The green grid represents the cone of influence that takes into account the edge effects and delimits the reliable rein in the WPS. Red and dark colours correspond to strong power while blue corresponds to weak power. Bottom upper panel: projection of the WPS on the time axis between periods of 1 and 4.1 days. Bottom lower panel: Temporal variation of the standard deviation for KIC~3733735 for subseries of length $5 \times P_{\rm rot}$. The black dashed line represents the mean value of the standard deviation of the subseries, i.e. the magnetic index $ \langle S_{\rm ph} \rangle$. The triple dot-dashed line represents the mean magnetic index for the low-activity subseries ($ \langle S_{\rm ph} (n_{low}) \rangle$) and the dash-dot line corresponds to the mean magnetic index for the high-activity subseries $ \langle S_{\rm ph} (n_{high}) \rangle$.}
\label{Kaa_wavelets}
\end{center}
\end{figure*}

In this work we use high-quality and long photometric data provided by the NASA {\it Kepler} satellite \citep{2010Sci...327..977B}. As we are only interested in studying periodic modulations in the light curves that are longer than a few hours, we are able to use long-cadence data \citep[29.43 minutes,][]{2010ApJ...713L.160G} that have been recorded since the beginning of the mission for most of the stars. Thus, 1440 days (Quarters 0 to 16) are used.  In Fig.~\ref{Kaa_wavelets} (top) 
 we show an example of a light curve for the {\it Kepler} target KIC~3733735. The light curves were corrected for instrumental effects and filtered with a 20-day triangular smooth \citep[][Mathur et al., in prep.]{2011MNRAS.414L...6G}.  To minimise the impact of wrong corrections in our data processing, we have also used PDC-msMAP light curves \citep{2012PASP..124.1000S, 2012PASP..124..985S}, as described in the Data Release Notes (DRN) 21 for Q0-Q14 and DRN20 and 21 for Q15  and Q16, respectively \citep{Thompson_2013, Thompson_2013b}. Although PDC-msMAP high-pass filters the light curves at periods below about 20 days, this does not affect our sample of stars because they have been selected to have rotation periods shorter than 12 days. Hence, we have verified that the retrieved rotation periods and the activity related modulations remained the same.


A sample of 196 solar-like stars were observed by {\it Kepler} since Q5 for at least three months in short cadence \citep[58.85 s, ][]{2010ApJ...713L.160G} as part of the {\it Kepler} Asteroseismic Science Consortium (KASC) program. Short-cadenced data are necessary to extract the seismic global properties of the stars in order to infer reliable stellar masses and radii as well as the structure of the stellar interior, in particular, the extension of their convective envelope.

 The surface rotation period of these stars at the active latitudes can be studied by looking at the photometric modulation induced by the star's active regions crossing  the stellar visible disk \citep[][Garc\'\i a et al. in prep.]{2009A&A...506..245M,2011A&A...530A..97B,2011A&A...534A...6C,2011ApJ...733...95M}. We expect that fast rotators have shorter cycle periods as shown from observations by \citet{2007ApJ...657..486B}. 
 Following this idea, we selected  22 stars that have rotation periods shorter than 12 days to look for magnetic activity cycles within the observation time of the {\it Kepler} mission ($\sim$~\,3.7 years so far). We focus on F stars with $T_{\rm eff} >$\,6000\,K, where the  $T_{\rm eff}$  values were taken from \citet{2012ApJS..199...30P}. F stars are interesting because they have thinner convective zone and present a strong increase in differential rotation \citep[e.g.][]{2012ApJ...756..169A}. Except the Sun, we note that so far the only star for which a cycle-like magnetic modulation has been detected using seismology is the F star HD~49933 \citep{2010Sci...329.1032G}. It is therefore suitable to concentrate our analysis on this subset of stars.

In Table~\ref{tbl-seismic} we present the seismic global parameters of the subset of 22 stars studied in this work.
They were obtained with the A2Z pipeline \citep{2010A&A...511A..46M} and are in agreement with the results from  \citet{2013arXiv1310.4001C} (listed as $M_{*}$ in Table~\ref{tbl-seismic}), 19 of them at a 1-$\sigma$ level. For the other three stars, the agreement in $\nu_{\rm max}$ is at a 2-$\sigma$ level.  A visual inspection of the 

\clearpage

\clearpage

\begin{sidewaystable*}[htdp]
\caption{List of the 22 stars analysed in this study with their  {\it Kepler} magnitude ($K_p$), their global seismic parameters, effective temperature from \citet{2012ApJS..199...30P}, $\log g$, mass, and radius obtained with the scaling relations, masses $M_{*}$ obtained with grid-based models from \citet{2013arXiv1310.4001C}, and surface rotation period as determined by Garcia et al. (in prep.).}
\begin{center}
\begin{tabular}{rrrrrcccccr}
\hline
\hline
{\bf Ref. no.} & KIC & $K_p$ & $\Delta \nu$ ($\mu$Hz) & $\nu_{\rm max}$ ($\mu$Hz)& $T_{\rm eff}$ (K) & $\log g$ & R ($R_{\odot}$) &M ($M_{\odot}$)& {\bf $M_{*}$ ($M_{\odot}$)} & $P_{\rm rot}$ (days)\\
\hline
 1& 1430163 &   9.58&  85.66\,$\pm$\, 1.80 &  1805.07\,$\pm$\,  29.66 &
6796\,$\pm$\, 78&4.25\,$\pm$\,0.01&1.59\,$\pm$\,0.10 &  1.63\,$\pm$\,0.25&
$1.38^{+0.09}_{-0.15}$& 3.88\,$\pm$\, 0.58 \\
 2& 1435467$^{a}$  &   8.88&  70.59\,$\pm$\, 1.41 &  1385.26\,$\pm$\,  72.04 &
6433\,$\pm$\, 86&4.12\,$\pm$\,0.03&1.75\,$\pm$\,0.17 &  1.47\,$\pm$\,0.38&
$1.19^{+0.08}_{-0.20}$& 6.53\,$\pm$\, 0.64 \\
 3& 3733735$^{a}$  &   8.37&  92.37\,$\pm$\, 1.72 &  2132.61\,$\pm$\,  84.44 &
6711\,$\pm$\, 66&4.32\,$\pm$\,0.02&1.61\,$\pm$\,0.14 &  1.95\,$\pm$\,0.42&
$1.36^{+0.12}_{-0.09}$& 2.54\,$\pm$\, 0.16 \\
 4& 4638884 &   9.86&  60.83\,$\pm$\, 1.15 &  1192.28\,$\pm$\,  75.82 &
6662\,$\pm$\, 57&4.06\,$\pm$\,0.03&2.07\,$\pm$\,0.23 &  1.79\,$\pm$\,0.52&
$1.47^{+0.11}_{-0.14}$& 6.17\,$\pm$\, 0.60 \\
 5& 5371516 &   8.37&  55.46\,$\pm$\, 1.06 &  1018.62\,$\pm$\,  59.84 &
6350\,$\pm$\, 81&3.98\,$\pm$\,0.03&2.07\,$\pm$\,0.22 &  1.51\,$\pm$\,0.42&
$1.38^{+0.13}_{-0.14}$& 5.13\,$\pm$\, 0.48 \\
 6& 5773345 &   9.16&  57.53\,$\pm$\, 1.26 &  1090.80\,$\pm$\,  63.97 &
6214\,$\pm$\, 61&4.01\,$\pm$\,0.03&2.04\,$\pm$\,0.23 &  1.55\,$\pm$\,0.45&
$1.49^{+0.13}_{-0.11}$&11.60\,$\pm$\, 0.90 \\
 7& 6508366$^{a}$  &   8.97&  51.83\,$\pm$\, 1.08 &   954.19\,$\pm$\,  58.57 &
6499\,$\pm$\, 46&3.96\,$\pm$\,0.03&2.25\,$\pm$\,0.25 &  1.68\,$\pm$\,0.49&
$1.40^{+0.11}_{-0.15}$& 3.78\,$\pm$\, 0.34 \\
 8& 7103006$^{a}$  &   8.86&  59.68\,$\pm$\, 1.35 &  1159.14\,$\pm$\,  49.55 &
6421\,$\pm$\, 51&4.04\,$\pm$\,0.02&2.05\,$\pm$\,0.20 &  1.68\,$\pm$\,0.41&
$1.37^{+0.15}_{-0.13}$& 4.61\,$\pm$\, 0.38 \\
 9& 7206837$^{a}$  &   9.77&  79.46\,$\pm$\, 1.59 &  1623.49\,$\pm$\,  91.07 &
6392\,$\pm$\, 59&4.19\,$\pm$\,0.03&1.62\,$\pm$\,0.17 &  1.46\,$\pm$\,0.40&
$1.29^{+0.13}_{-0.10}$& 4.04\,$\pm$\, 0.25 \\
10& 7668623 &   9.40&  46.29\,$\pm$\, 0.86 &   782.87\,$\pm$\,  36.85 &
6277\,$\pm$\, 48&3.87\,$\pm$\,0.02&2.28\,$\pm$\,0.21 &  1.39\,$\pm$\,0.33&
$1.36^{+0.16}_{-0.20}$& 5.12\,$\pm$\, 0.38 \\
11& 7771282 &  10.77&  72.32\,$\pm$\, 1.59 &  1435.17\,$\pm$\,  39.35 &
6407\,$\pm$\, 74&4.13\,$\pm$\,0.02&1.73\,$\pm$\,0.14 &  1.48\,$\pm$\,0.29&
$1.37^{+0.12}_{-0.10}$&11.91\,$\pm$\, 0.96 \\
12& 7940546 &   7.40&  58.87\,$\pm$\, 1.24 &  1077.78\,$\pm$\,  89.36 &
6350\,$\pm$\,111&4.01\,$\pm$\,0.04&1.95\,$\pm$\,0.26 &  1.41\,$\pm$\,0.50&
$1.32^{+0.11}_{-0.08}$&11.12\,$\pm$\, 0.85 \\
13& 8026226$^{a}$  &   8.43&  33.38\,$\pm$\, 1.42 &   560.74\,$\pm$\,  33.75 &
6319\,$\pm$\, 54&3.72\,$\pm$\,0.03&3.15\,$\pm$\,0.48 &  1.90\,$\pm$\,0.71&
$1.49^{+0.13}_{-0.15}$&11.14\,$\pm$\, 2.16 \\
14& 9226926 &   8.70&  73.77\,$\pm$\, 1.26 &  1411.43\,$\pm$\,   0.99 &
7149\,$\pm$\,132&4.15\,$\pm$\,0.00&1.72\,$\pm$\,0.07 &  1.53\,$\pm$\,0.14&
$1.41^{+0.09}_{-0.14}$& 2.17\,$\pm$\, 0.14 \\
15& 9289275 &   9.65&  66.92\,$\pm$\, 1.63 &  1379.79\,$\pm$\,  37.48 &
6227\,$\pm$\, 48&4.11\,$\pm$\,0.02&1.91\,$\pm$\,0.16 &  1.72\,$\pm$\,0.35&
$1.36^{+0.12}_{-0.08}$&11.31\,$\pm$\, 1.02 \\
16& 9812850$^{a}$  &   9.48&  65.03\,$\pm$\, 1.39 &  1231.57\,$\pm$\, 107.26 &
6407\,$\pm$\, 47&4.07\,$\pm$\,0.04&1.83\,$\pm$\,0.25 &  1.43\,$\pm$\,0.53&
$1.25^{+0.11}_{-0.16}$& 5.05\,$\pm$\, 0.53 \\
17& 10016239 &   9.91& 101.60\,$\pm$\, 2.59 &  2087.74\,$\pm$\,  22.83 &
6482\,$\pm$\, 51&4.30\,$\pm$\,0.01&1.28\,$\pm$\,0.09 &  1.19\,$\pm$\,0.19&
$1.28^{+0.11}_{-0.09}$& 4.92\,$\pm$\, 0.38 \\
18& 10162436$^{a}$  &   8.61&  55.82\,$\pm$\, 1.24 &  1006.55\,$\pm$\,  77.89 &
6346\,$\pm$\,108&3.98\,$\pm$\,0.04&2.02\,$\pm$\,0.26 &  1.42\,$\pm$\,0.49&
$1.28^{+0.12}_{-0.14}$&11.59\,$\pm$\, 0.88 \\
19& 10355856$^{a}$  &   9.19&  67.47\,$\pm$\, 1.43 &  1298.44\,$\pm$\,  64.19 &
6558\,$\pm$\, 56&4.10\,$\pm$\,0.03&1.82\,$\pm$\,0.18 &  1.50\,$\pm$\,0.38&
$1.39^{+0.08}_{-0.14}$& 4.47\,$\pm$\, 0.28 \\
20& 10644253$^{a}$  &   9.16& 123.04\,$\pm$\, 3.39 &  2924.64\,$\pm$\,  38.43 &
6069\,$\pm$\, 50&4.43\,$\pm$\,0.01&1.18\,$\pm$\,0.09 &  1.38\,$\pm$\,0.24&
$1.09^{+0.12}_{-0.10}$&10.89\,$\pm$\, 0.81 \\
21& 11070918 &   9.16&  66.89\,$\pm$\, 1.61 &  1093.64\,$\pm$\,  20.03 &
6720\,$\pm$\,  100&4.03\,$\pm$\,0.01&1.58\,$\pm$\,0.12 &  0.96\,$\pm$\,0.17&
$1.32^{+0.17}_{-0.11}$& 2.91\,$\pm$\, 0.18 \\
22& 12009504$^{a}$  &   9.32&  88.34\,$\pm$\, 2.27 &  1848.45\,$\pm$\, 133.86 &
6270\,$\pm$\, 61&4.24\,$\pm$\,0.04&1.47\,$\pm$\,0.19 &  1.37\,$\pm$\,0.47&
$1.25^{+0.10}_{-0.09}$& 9.48\,$\pm$\, 0.85 \\

\hline
\end{tabular}
\end{center}
\label{tbl-seismic}
\tablebib{(a) Analysed in \citet{2012A&A...537A.134A}.}
\end{sidewaystable*}%

\noindent  power spectra and the global seismic parameters of these stars showed that the disagreements can be explained. The difference between the two analyses can be attributed to the longer datasets we are using here compared to the \citet{2013arXiv1310.4001C} datasets, which are based on 1-month {\it Kepler} data from the survey phase. 
Moreover, KIC~9289275 and 1016239 have what is known as a double bump acoustic (p)-mode envelope that complicates the determination of $\nu_{\rm max}$ \citep{2010ApJ...713..935B}. 

\citet{2012A&A...543A..54A} have already extracted the p-mode frequencies for 11 stars in our sample (indicated in Table~\ref{tbl-seismic}). The complete study of the individual p modes of these stars is out of  the scope of this paper. For our purposes, we can compute their masses and radii relying on the scaling relations as defined by \citet{kjeldsen95,2011A&A...529L...8K} and validated later by different methods \citep[e.g.][]{2011ApJ...743..143H,2012ApJ...749..152M,2012ApJ...757...99S}. The values that we obtain agree within 1-\,$\sigma$ (except KIC~3733375 which is within 2-\,$\sigma$) with the grid-based modelling results presented in Chaplin et al. (submitted). We note that we used the observed mean large separation in the scaling relations, but the use of the asymptotic mean large separation as described in \citet{2013A&A...550A.126M} leads to smaller radii and masses, which also agree within the uncertainties with the values listed in Table~1.

The rotation periods obtained by Garc\'ia et al. (in prep.) are also given in Table~\ref{tbl-seismic}  with the associated uncertainties.  
Briefly, to extract the rotation periods we apply two different techniques to the two above-mentioned datasets. On the one hand, we perform a time-frequency analysis of the light curves using a Morlet transform and we project the result on the frequency domain \citep[e.g.][]{2010A&A...518A..53M,2010A&A...511A..46M,2013A&A...550A..32M}. In most of the cases, the rotation period would be the highest peak in this periodogram. Sometimes, when there are two active longitudes in each side of the star, the highest peak retrieved is not the rotation period but the second harmonic. To avoid this case we also extract the second highest peak in the periodogram and we check if they correspond to a harmonic. On the other hand, we compute the autocorrelation of the temporal signal to extract a second estimation of the rotation rate. The autocorrelation is a powerful method as has been demonstrated by \citet{2013MNRAS.432.1203M}. The use of both algorithms, wavelets and autocorrelation, allows us to minimise any false detection or ambiguities among the harmonics of the signal.

We also checked the stability of the autocorrelation function to study the lifetime of the starspots or active regions (at active longitudes) crossing the visible disk of the star. \citet{2013MNRAS.432.1203M} used the stability of the autocorrelation function to point out possible classical pulsators for which low-frequency modes are very stable, but this stability can also be due to long-lived active regions or stable groups of starspots at a given active longitude on the star \citep[e.g.][]{2011A&A...532A..81F,2009IAUS..259..363S}. These long-lived structures, if they are rotating at close rates (due to differential rotation), can produce a beating in the light curve that mimics the effect of activity cycles.

\begin{figure}[htbp]
\begin{center}
\includegraphics[angle=90, width=9.5cm]{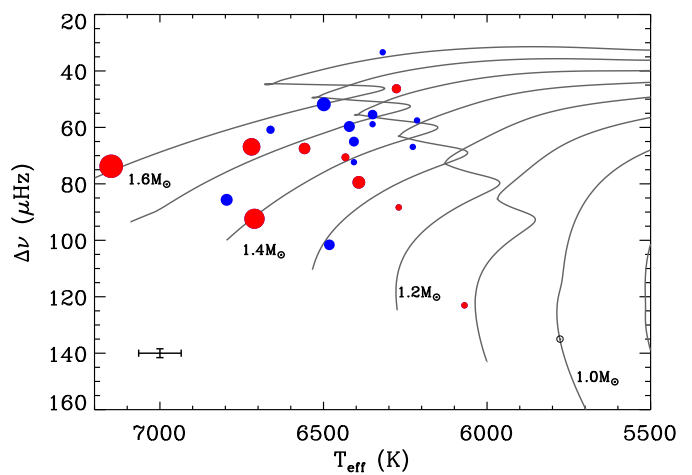}
\caption{ Modified HR diagram showing the sample of 22 F stars analysed. Red symbols represent the stars that are discussed in Section 4.2: with long-lived features, with cycle-like behaviour, or with a trend in the magnetic proxy. Blue symbols correspond to stars where no clear cycle is observed for the duration of the observations also referred as the ``Other'' category in Table~2. The size of the symbols is inversely proportional to the surface rotation period of the star. The position of the Sun is indicated by the $\odot$ symbol, and evolution tracks from Aarhus Stellar Evolution Code (ASTEC) are shown for a range of masses at solar composition ($Z_\odot$ = 0.0246). Median uncertainties on $\langle \Delta \nu \rangle$ and $T_{\rm eff}$ are shown in the lower left corner of the figure.}
\label{HRdiag}
\end{center}
\end{figure}

In Fig.~\ref{HRdiag}, we show a seismic Hertzsprung-Russell (H-R) diagram with the 22 stars in our sample, where we used the mean large frequency separation ($<\Delta \nu>$) instead of the luminosity on the vertical axis. There could be a small difference between the mass derived by \citet{2013arXiv1310.4001C} and the position of the star on the HR diagram that could be due to differences in metallicity.



In order to verify that the modulations observed in the light curves were genuine and not due to pollution from nearby stars (active or classical pulsators), the pixel data of each star were examined. We first checked the full frame image of the corresponding channels provided by the Mikulski Archive for Space Telescopes (MAST) to ensure that the considered stars were not included in the halo of a nearby pulsating star, which could account for a false modulation of the obtained light curve. This was not the case for any of the stars in our study. We then performed a more thorough verification using the pixel data itself (see Mathur et al. in prep.). For each star, we divided the selected stellar mask into several areas, typically four quadrants, and looked for the modulation in the light curves extracted from these sub-masks. The presence of a nearby pulsating star in the mask of the observed star could have led to the presence of a modulation of the signal in a limited area of the mask. As this was not the case for the stars studied in this paper, we can reasonably consider that the observed modulations of their light curves are intrinsic.


\section{Magnetic index}

It would be interesting to compute a magnetic index to quantify the level of magnetic activity of different stars and be able to compare them. To do so, we need to define a magnetic index --similar to the average Mount Wilson S-index \citep{1982A&A...107...31M}-- based on photometric observations instead of chromospheric ones. While studying the CoRoT star HD~49933, \citet{2010Sci...329.1032G} defined a magnetic index that is the standard deviation of 30-day-long subseries in the light curve, yielding a magnetic-cycle proxy computed as the temporal variation of this standard deviation. This calculation was later taken by \citet{2011ApJ...732L...5C} to study the impact of stellar magnetic activity on the acoustic modes and their detectability. They slightly modified the computation of the index by measuring the standard deviation of the {\it Kepler} light curves that were low-pass filtered and smoothed by a 1 hr boxcar.

In this paper, we propose to use the standard deviation of the whole time series as a magnetic index of each star. We call this index $S_{\rm ph}$ as a counterpart of the spectroscopic S-index. It is sensitive to the stellar variability but also to the photon noise $\sigma_{\rm ph}$. There are several ways of estimating the contribution of $\sigma_{\rm ph}$ in the {\it Kepler} light curves. Here we used the approach proposed by \citet{2010ApJ...713L.120J}. They used a sample of 1639 stars obtained with one month of {\it Kepler} data in long cadence to infer an empirical relation between the apparent magnitude of the stars and $\sigma_{\rm ph}$. This relation was then successfully used by \citet{2011ApJ...741..119M} to characterise the long-cadence background of more than a thousand {\it Kepler} red giants. We note that this relation depends on the magnitude of the stars and on the way the light curves are extracted (e.g. the size of the masks used in each star). 


 As said before, $S_{\rm ph}$ is obtained for the whole time series. This allows us to minimise the effects of any instrumental glitch in the data. However, $S_{\rm ph}$ is sensitive to the presence or to the absence of magnetic cycles during the observations, which can bias the magnetic index (as happens with the S-index).To assess the existence of variations in the activity levels in a given star we also compute $S_{\rm ph}$ locally, during low and high activity periods, to determine the ratio or contrast between periods with different average activity. To do so, we first measure the standard deviation for independent subseries of length $k \times P_{\rm rot}$, where $k$ is fixed to 5. Indeed, we have verified that when $k > 3$, the standard deviation converges towards a constant value ensuring the universal character of this index.
 Then we select the subseries where the standard deviation is smaller ($n_{low}$) or greater ($n_{high}$) than $S_{\rm ph}$. We also define another global index $ \langle S_{\rm ph} \rangle$ as the mean value of all the subseries we have, after removing the null values. This parameter is very close to $S_{\rm ph}$. Figure~\ref{Kaa_wavelets} shows an example of the sigma computed for the different subseries for one star in our sample: KIC~3733735. The dashed line represents the mean value of the standard deviation $ \langle S_{\rm ph} \rangle$. We also note that these indexes are sensitive to the inclination angle of the star. The ratio
\begin{equation}
C =  \langle S_{\rm ph} (n_{high}) \rangle /  \langle S_{\rm ph} (n_{low}) \rangle
\end{equation}

\noindent
gives a value of the contrast between periods of high activity and low activity during the observations. This is a useful way to disentangle stars potentially showing a magnetic activity cycle during the observations.

 In Table 2 the global $\langle S_{\rm ph} \rangle$ index corrected for the photon noise as computed above, as well as the local indexes, also corrected for the photon noise: $ \langle S_{\rm ph} (n_{low}) \rangle$ and $ \langle S_{\rm ph} (n_{high}) \rangle$, are given for the 22 F stars and the Sun using subseries of length $5 \times P_{\rm rot}$. To compute the solar magnetic index $<S_{\rm ph,\odot}>$ we used the photometric data recorded by the Sun Photometers (SPM), belonging to the VIRGO \citep[Variability of solar Irradiance and Gravity Oscillations;][]{1995SoPh..162..101F} instrument on board the Solar and Heliospheric Observatory (SoHO). We summed the two channels (green and red) to be close to the {\it Kepler} bandwidth and rebinned the data so that they have a sampling of 30 minutes like the {\it Kepler} long-cadence data. We obtained  a value of $<S_{ph,\odot}>$= 166.1 ppm  for the mean activity level of the Sun integrated over the last 16 years (most of cycle 23 and the rising phase of cycle 24) \citep[see][for more details]{2013JPhCS.440a2020G}. Finally, we computed the associated statistical uncertainties $\epsilon$ by taking the dispersion between the different subseries used for the calculation of $ \langle S_{\rm ph} \rangle$.
 
\begin{figure}[htbp]
\begin{center}
\includegraphics[angle=90, width=9cm]{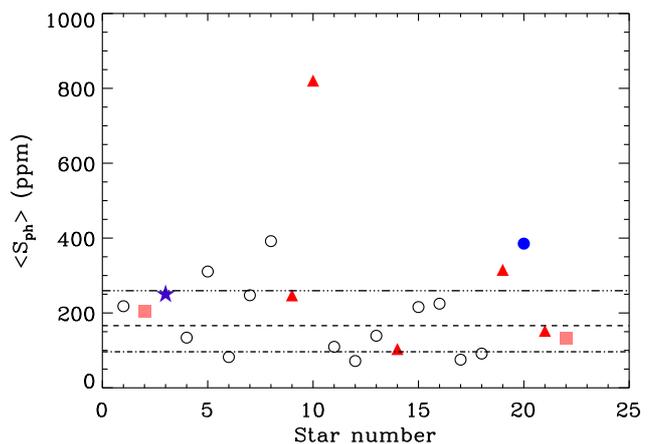}
\caption{Mean magnetic index $ \langle S_{\rm ph} \rangle$ identified by the star number given in Table 1. The stars are represented with different symbols according to their category in Table~2: L (red triangles), C (blue filled circle), T (pink squares), and O (open circles). KIC~3733735 is represented with the blue star since it belongs to both categories L and C. We superimposed the mean index of the Sun  $<S_{\rm ph, \odot}>$=166.1\,$\pm$\,2.6\,ppm (dashed line), the index at low activity $<S_{\rm ph}(n_{\rm low})>$, (dot dash line), and the index at high activity $<S_{\rm ph}(n_{\rm high})>$ (triple dot dash line) obtained with the VIRGO data.}
\label{Sphotom_22stars}
\end{center}
\end{figure}

 Figure~\ref{Sphotom_22stars} shows the mean activity index $ \langle S_{\rm ph} \rangle$ for the 22 stars and the Sun. We notice that $ \langle S_{\rm ph} \rangle$ varies between 70 to 900~ppm with an average activity level for our sample of stars at around 220~ppm showing that the magnetic activity of our set of stars is in general slightly larger than the mean activity level of the Sun.
 This is not in agreement with the results by \citet{1995ApJ...438..269B} who concluded that the F stars observed by the Mount Wilson survey had low S-indexes with a rather flat behaviour or a long-term variability, which they attributed to the lower chromospheric activity of these stars. Our analysis suggests that the photospheric activity is higher compared to the Sun. However, we have not taken into account any projection effects due to lower inclination stellar angles.  Finally, KIC~7668623 has the largest value of the magnetic index. This case will be further discussed in Section 4.2.

\begin{figure}[htbp]
\begin{center}
\includegraphics[angle=90, width=9cm]{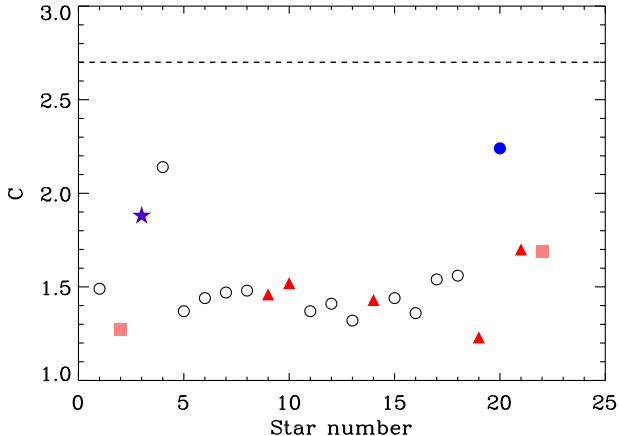}
\caption{Contrast, $C$, between high and low magnetic activity for the 22 F stars for $k$=5. The dashed line represents the contrast for the Sun obtained with the VIRGO data. Same legend as in Figure 3.}
\label{C_22stars}
\end{center}
\end{figure}

We also show the contrast, $C$, between the maximum and minimum activity (Fig.~\ref{C_22stars}). While most of the stars lie between 1 and 2, the Sun stands higher at 2.5. Only one star is close to the Sun, KIC~10644253, and we will come back to it in the discussion. 


Finally, we compute the range or variability index $R_{\rm var}$ defined by \citet{2011AJ....141...20B,2013ApJ...769...37B}. They take subseries of 30 days and compute this index by ranking all the points of the light curve and removing the 5\% of the points with the lowest and highest flux values. Then they measure the difference between the highest flux and the lowest flux remaining, which gives the range of flux spanned by the time series.  The final index is the median value of all the subseries. Because this analysis is done for subseries of 30 days, it assumes that the rotation period of the star is shorter than 30 days, otherwise we cannot follow the change of the active regions. Fortunately for our stars, the rotation periods are below 12 days. The values of $R_{\rm var}$ are given in Table 2 and they are also corrected for the photon noise. We notice that some stars with large values of $ \langle S_{\rm ph} \rangle$ do not have large values of $R_{\rm var}$. This is an inherent consequence of the definition of $R_{\rm var}$. Assuming a proper correction of the light curve, the truncation of the highest 5\% of the signal removes the highest activity points and results in a lower index relative to $ \langle S_{\rm ph} \rangle$.


\begin{table*}[htdp]
\caption{Different indexes obtained for the 22 stars: the standard deviation of the whole time series ($S_{\rm ph}$), the magnetic index corrected from the photon noise as defined in Section~3 ($ \langle S_{\rm ph} \rangle$), the magnetic index at high activity ($\langle S_{\rm ph}(n_{\rm high}) \rangle$), the magnetic index at low activity ($\langle S_{\rm ph}(n_{\rm low}) \rangle$), the contrast (C), and the range ($R_{\rm var}$) as defined by \citet{2011AJ....141...20B} and corrected for the photon noise. }
\begin{center}
\begin{tabular}{rcccccccccc}
\hline
\hline
Ref. no. & KIC & $S_{\rm ph}$ (ppm)& $ \langle S_{\rm ph} \rangle$ (ppm)& $\langle S_{\rm ph}(n_{\rm high}) \rangle$ (ppm)&  $\langle S_{\rm ph}(n_{\rm low}) \rangle$ (ppm)& C & $R_{\rm var}$ (ppm) & Category$^a$\\
\hline
 1& 1430163 &   231.7 &   218.0\,$\pm$\,   8.4& 268.1\,$\pm$\,  10.2 & 
 179.8\,$\pm$\,  10.2 &  1.5 &  498.5&O\\
 2& 1435467 &   211.0 &   205.5\,$\pm$\,   5.8& 226.0\,$\pm$\,   6.3 & 
 178.6\,$\pm$\,   6.3 &  1.3 &  506.7&T\\
 3& 3733735 &   271.8 &   249.9\,$\pm$\,  11.9& 343.9\,$\pm$\,  15.3 & 
 183.0\,$\pm$\,  15.3 &  1.9 &  625.9&L+C\\
 4& 4638884 &   148.7 &   134.1\,$\pm$\,   4.5& 213.1\,$\pm$\,   6.4 & 
  99.5\,$\pm$\,   6.4 &  2.1 &  281.0&O\\
 5& 5371516 &   321.6 &   310.7\,$\pm$\,  10.6& 361.4\,$\pm$\,  11.7 & 
 264.0\,$\pm$\,  11.7 &  1.4 &  581.8&O\\
 6& 5773345 &    85.9 &    82.2\,$\pm$\,   2.3& 102.2\,$\pm$\,   2.7 & 
  70.8\,$\pm$\,   2.7 &  1.4 &  167.8&O\\
 7& 6508366 &   259.1 &   247.3\,$\pm$\,   9.3& 310.2\,$\pm$\,  11.3 & 
 210.6\,$\pm$\,  11.3 &  1.5 &  402.0&O\\
 8& 7103006 &   406.6 &   391.9\,$\pm$\,  13.1& 470.5\,$\pm$\,  15.5 & 
 318.1\,$\pm$\,  15.5 &  1.5 &  672.2&O\\
 9& 7206837 &   255.8 &   247.1\,$\pm$\,   9.1& 302.2\,$\pm$\,  10.9 & 
 206.8\,$\pm$\,  10.9 &  1.5 &  438.3&L\\
10& 7668623 &   851.9 &   820.7\,$\pm$\,  25.6&1020.2\,$\pm$\,  30.6 & 
 671.0\,$\pm$\,  30.6 &  1.5 & 1630.2&L\\
11& 7771282 &   112.0 &   109.5\,$\pm$\,   2.8& 128.1\,$\pm$\,   3.1 & 
  93.7\,$\pm$\,   3.1 &  1.4 &  210.7&O\\
12& 7940546 &    74.8 &    71.7\,$\pm$\,   1.6&  86.8\,$\pm$\,   1.9 & 
  61.6\,$\pm$\,   1.9 &  1.4 &  180.5&O\\
13& 8026226 &   141.8 &   139.1\,$\pm$\,   3.0& 164.2\,$\pm$\,   3.5 & 
 124.9\,$\pm$\,   3.5 &  1.3 &  265.9&O\\
14& 9226926 &   108.9 &   103.7\,$\pm$\,   5.4& 124.8\,$\pm$\,   6.2 & 
  87.3\,$\pm$\,   6.2 &  1.4 &  238.3&L\\
15& 9289275 &   222.8 &   215.8\,$\pm$\,   4.8& 260.1\,$\pm$\,   5.6 & 
 180.9\,$\pm$\,   5.6 &  1.4 &  452.6&O\\
16& 9812850 &   228.7 &   224.7\,$\pm$\,   7.4& 260.3\,$\pm$\,   8.4 & 
 191.6\,$\pm$\,   8.4 &  1.4 &  526.7&O\\
17& 10016239 &    78.6 &    75.2\,$\pm$\,   3.0&  94.8\,$\pm$\,   3.5 & 
  61.4\,$\pm$\,   3.5 &  1.5 &  149.9&O\\
18& 10162436 &    95.2 &    91.5\,$\pm$\,   2.5& 111.4\,$\pm$\,   3.0 & 
  71.6\,$\pm$\,   3.0 &  1.6 &  225.6&O\\
19& 10355856 &   321.0 &   315.2\,$\pm$\,  11.2& 354.5\,$\pm$\,  13.1 & 
 287.3\,$\pm$\,  13.1 &  1.2 &  683.5&L\\
20& 10644253 &   433.4 &   385.1\,$\pm$\,   8.9& 625.1\,$\pm$\,  13.1 & 
 278.5\,$\pm$\,  13.1 &  2.2 &  507.8&C\\
21& 11070918 &   162.9 &   152.2\,$\pm$\,   7.0& 196.8\,$\pm$\,   8.6 & 
 115.9\,$\pm$\,   8.6 &  1.7 &  210.3&L\\
22& 12009504 &   137.6 &   131.2\,$\pm$\,   3.4& 191.2\,$\pm$\,   4.7 & 
 113.0\,$\pm$\,   4.7 &  1.7 &  210.0&T\\

&Sun (G+R) & 192.2 & 166.1\,$\pm$\,2.6& 258.5\,$\pm$\,3.5 & 89.0\,$\pm$\,1.5 & 2.7 & 296.7 & C\\

\hline
\end{tabular}
\end{center}
\label{Tbl2}
\tablefoot{ \tablefoottext{a} C = Cycle; L=Long-lived spots/active regions; T=Trend; O = Others . See Section 4.2 for explanation.}
\end{table*}%

\section{Time-frequency analysis: Magnetic cycle proxy}

It is also interesting to study the temporal evolution of the magnetic activity of the stars and look for signatures of any cycle-like behaviour. To perform a time-frequency analysis, we used the wavelets tool developed by \citet{1998BAMS...79...61T}, including the correction by \citet{liu2007}, and adapted for asteroseismology by \citet{2010A&A...511A..46M}. It consists of looking for the correlation between the time series and a mother wavelet with a given period. We chose the Morlet wavelet, which is the convolution between a Gaussian function and a sinusoid. Then we slide the wavelet along time in the time series and we vary the period in a given range. This produces the wavelet power spectrum (WPS) represented in Figure~\ref{Kaa_wavelets} (middle panel). Given the stellar rotation period, we project the WPS on the time axis in a small region around this period. This leads to the scale average variance (SAV, bottom panel in Figure~\ref{Kaa_wavelets}), which corresponds to a magnetic proxy of any magnetic activity cycle. This ensures that the magnetic proxy is only related to structures rotating with the rotation period of the star and is not affected by a localised problem due to a bad correction of the {\it Kepler} data.

\subsection{The Sun}

For this analysis, we used the VIRGO/SPM data again. \citet{2013JPhCS.440a2020G} applied the same methodology that we have used in this work to 16 years of solar data measured by SoHO.  In Fig.~5 of this paper the scale average variance was compared to a typical solar magnetic activity proxy, the solar radio flux at 10.7 cm. 
We also note in their plot that during the rising phases of cycles 23 and 24 and during the maximum of cycle 23, some spikes appear in the magnetic proxy with a much higher amplitude compared to the minimum magnetic activity period. These features were also present in the solar radio flux, reinforcing the idea that the scale average variance is a good magnetic activity proxy at all temporal scales.

\begin{figure*}[htbp]
\begin{center}
\includegraphics[trim=0 0 4cm 6cm, width=6cm]{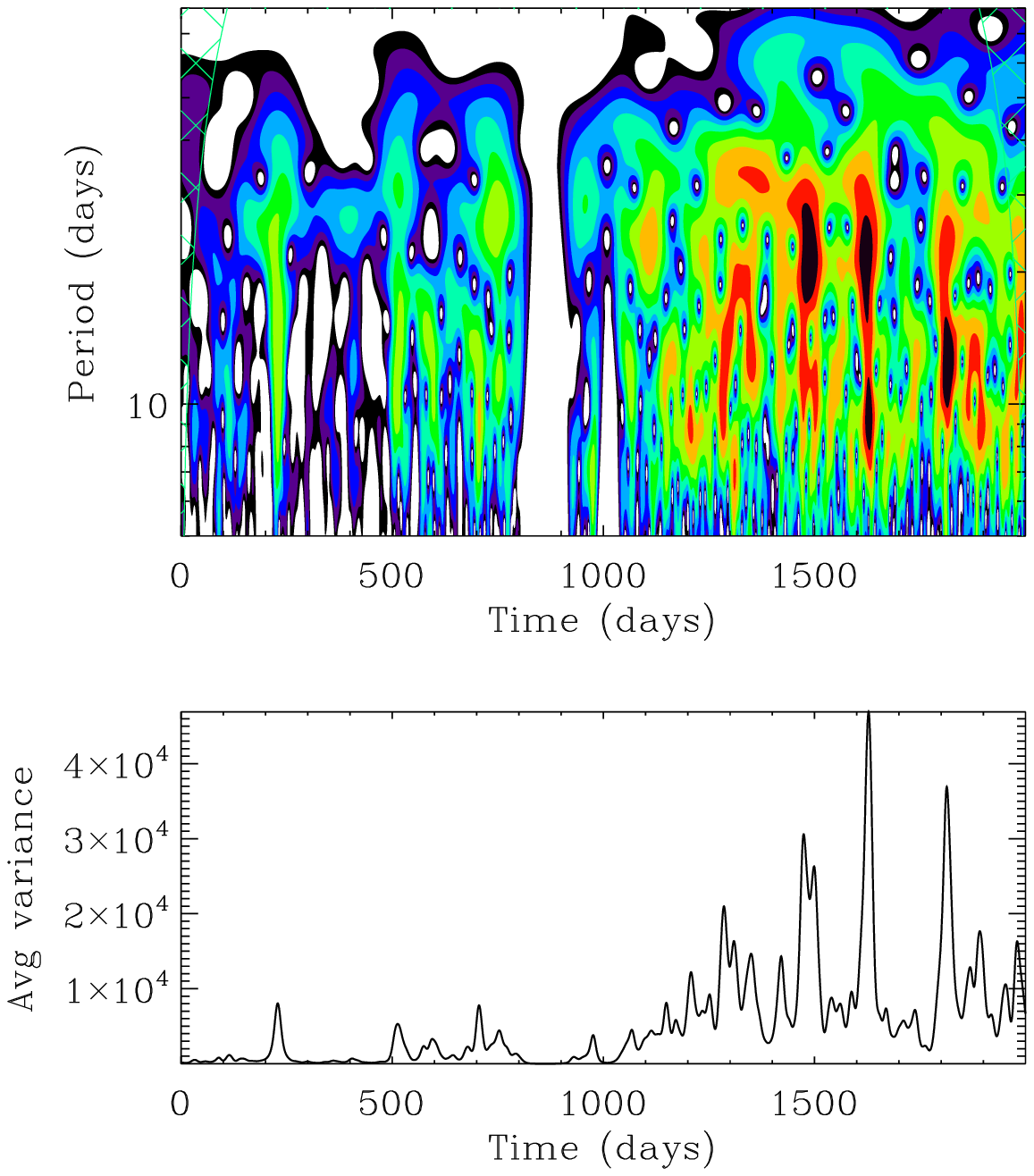}
\includegraphics[trim=0 0 4cm 6cm, width=6cm]{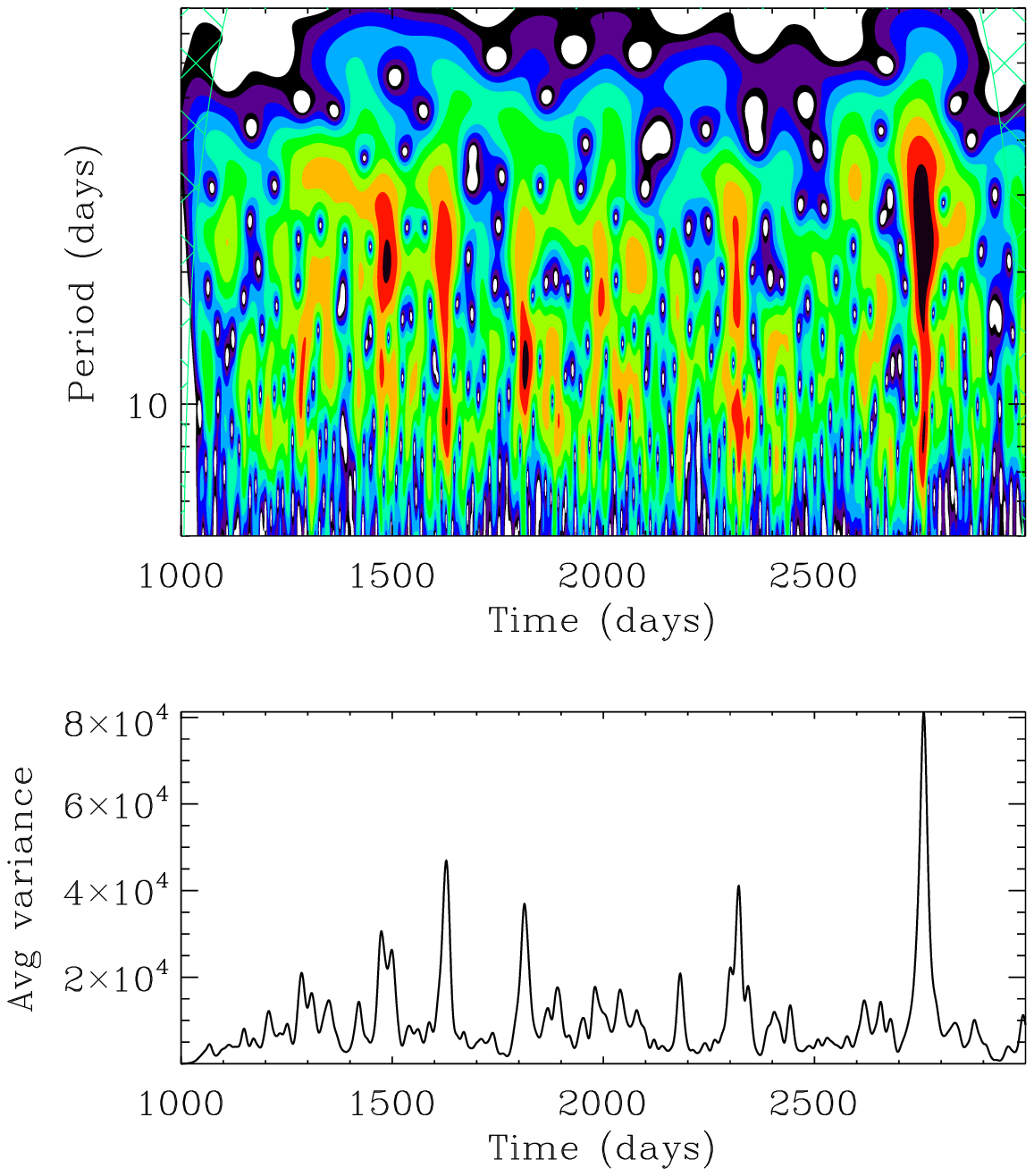}\\
\includegraphics[trim=0 0 4cm 6cm, width=6cm]{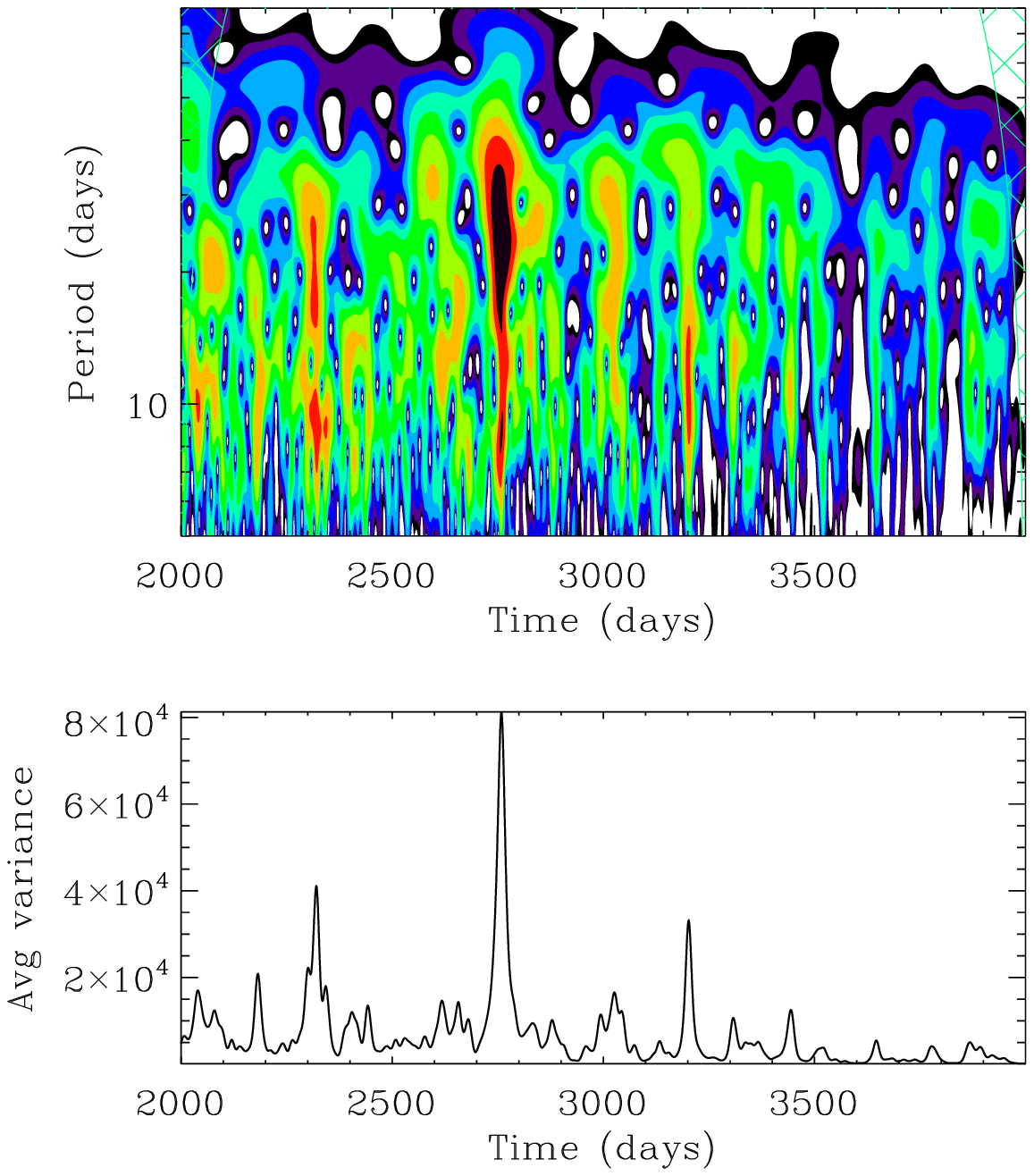}
\includegraphics[trim=0 0 4cm 6cm, width=6cm]{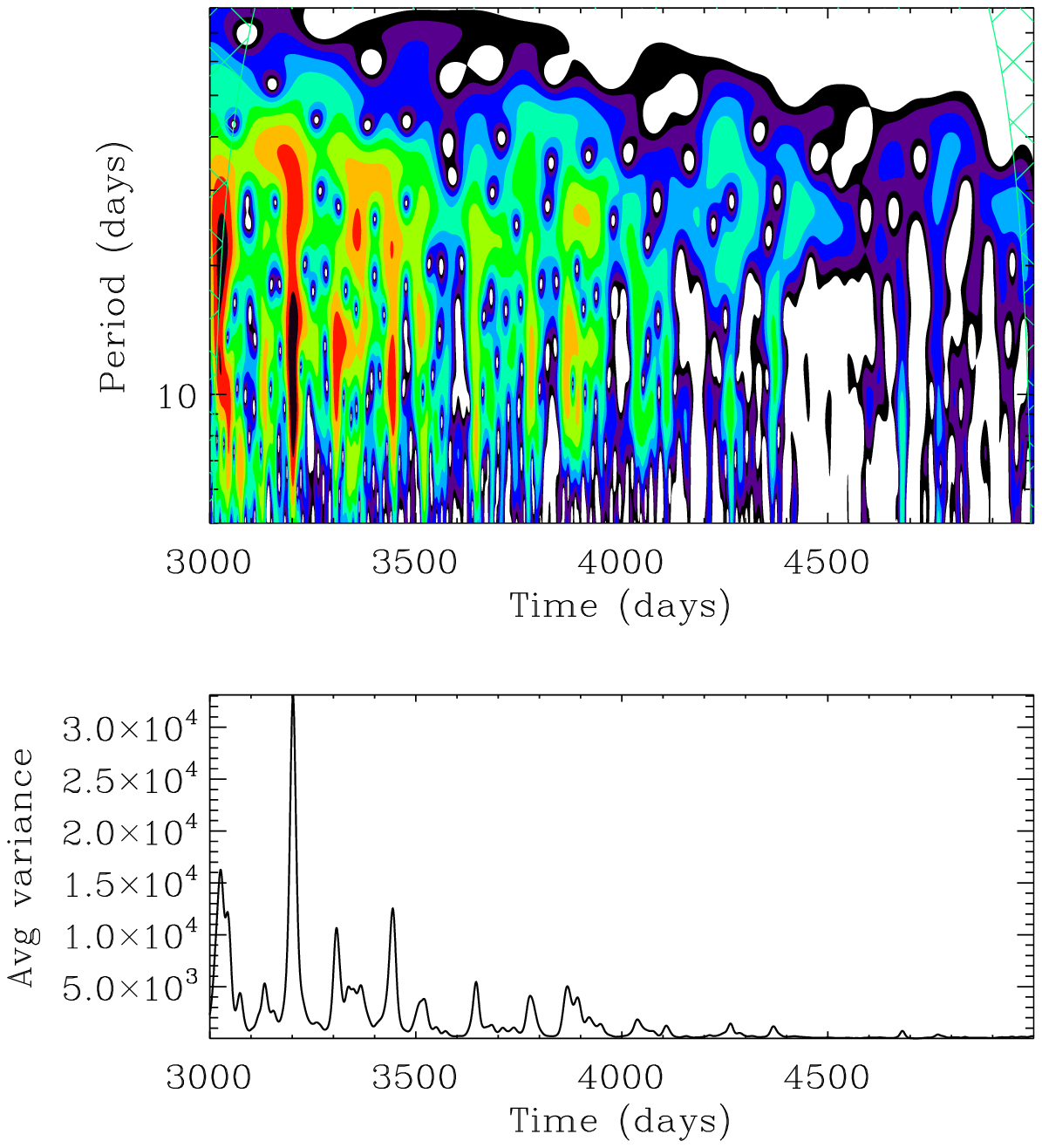}

\caption{ Wavelet analysis of VIRGO data for subseries of 2000 days. For each subseries, we show the wavelet power spectrum (red and dark areas are correspond to high power) and the average variance that is the magnetic proxy obtained by projecting the WPS between 6- and 60-day periods on the x-axis. Plots from top left to bottom right correspond to the increase in cycle 23 to the decrease in cycle 23.}
\label{Sun_wavelets}
\end{center}
\end{figure*}

From the {\it Kepler} observations, we do not expect to observe many complete activity cycles. This might be possible only for a handful of stars among the fastest rotators. We investigate here if the SAV is able to disentangle the different phases of a cycle when the observations span only part of it. This analysis will also allow us to characterise features that could help us later to identify at which moment of the magnetic cycle the stars are observed when only a portion of their activity cycles is sampled.


To better interpret the results of the wavelet analysis for our sample of F stars, we divided the VIRGO/SPM data into chunks of 2000 days starting on 11 April 1996. This corresponds to an observation length that is 70 times the equatorial rotation period of the Sun, which is $\sim$26 days. This is the case for our sample of {\it Kepler} stars, for which the rotation periods are shorter than 12 days ,and for which we have observed at least 70 rotation periods. The different subseries that we analysed are representative of different periods in the magnetic activity cycle of the Sun: the rising phase of cycle 23, the maximum activity of this cycle, its decrease, and the extended minimum between cycles 23 and 24. For the following study, the  scale average variance was obtained projecting all the wavelets between 6 and 60 days to be very conservative and to include the second harmonic or the rotation period of $\sim$13 days.

The top left panel in Fig.~\ref{Sun_wavelets} corresponds to the increase in activity for cycle 23. We see that in the first 1000 days the magnetic proxy is very flat. After that flat period, the magnetic proxy increases, so the wavelet analysis is able to see the rise in the magnetic cycle. 

The analysis of the period between 1000 and 3000 days, which includes the rising phase and the maximum of the magnetic activity cycle, is shown in the top right panel in Fig.~\ref{Sun_wavelets}. We notice that the magnetic proxy has quite a few high peaks whose amplitudes increase slightly. But having only these 2000 days of data at this given moment would not have allowed us to determine in which exact period of the cycle we are observing the Sun, except that it is quite active. 

The bottom left panel in Fig.~\ref{Sun_wavelets} represents the analysis of the maximum of cycle 23 and the beginning of the decrease of the cycle between 2000 and 4000 days. The magnetic proxy is flatter than the previous phase with amplitude peaks that are not as high. In particular, the last 800 days are flat and very similar to the first 1000 days of data. 

Finally, the bottom right panel in Fig.~\ref{Sun_wavelets} corresponds to the decrease of cycle 23 and the extended minimum between cycles 23 and 24. A few peaks are still present during the decrease of the cycle but after 4000 days the magnetic proxy is even flatter than the top left panel of the figure, so our magnetic proxy does see the extended minimum.


From this thorough analysis, we can say that the magnetic proxy SAV allows us to flag a rise or decrease in a magnetic activity cycle and therefore to uncover a cycle-like behaviour. However, if we observe most of the maximum activity cycle, it is very complicated to determine if the star is close to the rise or the dawn of the cycle. This is where the magnetic index $S_{\rm ph}$ intervenes and can help to disentangle in which situation the star is observed. 

We note that this analysis is done for the Sun, which has an inclination angle of 90$^\circ$, but this will not be the case for all our stars. We will come back to this problem in Section 5.2.

\subsection{The Kepler F stars}

We performed a time-frequency analysis by applying the wavelets to the 22 stars in our sample (see Fig.~\ref{Kaa_wavelets} and Appendix A)  yielding their magnetic activity proxy as defined above. The projection on the time axis was computed in a range of periods that takes into account the power seen in the WPS. The range used for each star is given in the caption of the figures. We notice different patterns that could correspond to different magnetic behaviour that is intrinsic in the stars and we divide our sample into four groups: stars with long-lived features (probably spots or active regions), stars with a cycle-like behaviour, stars that show an increase or decrease in the magnetic proxy, and stars that do not show any clear modulation in the magnetic proxy.

The first group consists of six stars: KIC~3733735, KIC~7206837, KIC~7668623, KIC~9226926, KIC~10355856, and KIC~11070918. We call this group `L' for long-lived features. For these six stars, we note that their WPS have a high level of power throughout almost all of the observation period. This suggests that there are spots or active regions that are present around the same region (or longitude) at the surface of the stars for a long period of time. 
Two of these stars, KIC~3733735 and KIC~9226926, have their magnetic proxies that  increase and decrease with time. This could be due to either an increase in the number of spots (similar to what we observe for the Sun) or a beating between two close frequencies that are very stable with time (active regions or modes). This will be developed further in Section 5.1. In addition, KIC~7206837, KIC~7668623, KIC~10355856, and KIC~11070918 have strong power in the second and sometimes third harmonics of the main period of rotation. This is probably due to the existence of several long-lived active regions at given longitudes on the surface of the stars. In particular, the magnetic proxy of KIC~9226926 shows a modulation of about 540 days on average while KIC~10355856 has a modulation of approximately 30 days in the magnetic proxy. We also note that KIC~7668623 has some large peaks in the SAV, which could be related to a low inclination angle of the star.

 The case of KIC~3733735 is very interesting because we observe two different modulations. The one of $\sim$\,90 days seems very similar to the previous stars suggesting the presence of stable active regions. This modulation appears to be triggered by the other modulation, which corresponds to a flat magnetic proxy during the first 600 days of observations followed by a progressive increase and decrease. This large envelope is very similar to what we observe in the Sun during one cycle giving us a hint that one magnetic cycle is observed for this star. 

The second group of stars (listed as C in Table 2) corresponds to those with cycle-like behaviour: KIC~3733735 (also in the previous category of long-lived spots) and KIC~10644253. The second star shows a few large peaks in the magnetic proxy suggesting that this star is observed with a low inclination angle and for which two magnetic cycles could have been observed.



The third group (listed as `T' in Table~2) contains two stars that show some trend in the magnetic activity proxy. KIC~1435467 seems to show a slight increase in the magnetic proxy.   The SAV of KIC~12009504 has a slight decrease and then an increase in the magnetic proxy. This might correspond to the observation of the transition from one cycle to the following one. Unfortunately, since it is difficult to interpret these trends when we do not see any flat behaviour, more data are needed to confirm these conclusions.

The fourth and last group contains 13 stars (category `O' in Table~2) that do not show any modulation that looks like a cycle. The magnetic proxy of these stars is more chaotic. A few simple explanations can be given for these stars: 1) the cycle period is longer than the observation length preventing us from observing a cycle; 2) these stars do not have regular cycles; or 3) their inclination angles are low. 


Since the magnetic proxy depends on the range of periods where we project the WPS, we cannot directly compare the absolute values of our stars with our study of the Sun. However, we can still  compare the global behaviour of the proxy.

The stars KIC~1430163, 5371516, 5773345, 7103006, 9812850, and 10016239 show a large number of peaks in the magnetic proxy as well as an almost continuous power in the wavelet power spectrum. This suggests that these stars have active regions that evolve, appear, and disappear with time. This looks very similar to the analysis of the VIRGO data during the rise and maximum of the cycle (top right panel in Fig.~\ref{Sun_wavelets}). As said before, these stars could be in a non-cycling state or being observed during a maximum of a much longer activity cycle. 

We note that KIC~4638884 has large peaks in the magnetic proxy. This  could be related to the inclination angle of the stars and the position of the active latitudes. We will discuss theses cases further in the next section.

We find six stars (KIC 6508366, 7771282, 7940546, 8026226, 9289275, and 10162436) that have very erratic behaviour preventing us from confirming any cycle or magnetic-related cycles detection.

\section{Discussion}

\subsection{Long-lived features}

In Section 4.2, we mentioned two stars, KIC~3733735 and KIC~9226926, where we see an interesting modulation in the magnetic proxy that could be due to the magnetic activity cycle of the stars.  One of these stars has a rather high magnetic index $\langle S_{\rm ph} \rangle$ of  249.9\,ppm while the other star has a rather low value of 103.7\,ppm. We looked in more detail into the low-frequency part of their power spectra around the rotation period found. We noticed that these stars have two or three high-amplitude peaks very close to each other. The existence of two close frequencies, $f_1$ and $f_2$, leads to a beating effect generating a sine wave of frequency $f_1 + f_2$ modulated by an envelope with a frequency $f_2-f_1$. 

For KIC~3733735, the two highest peaks are located at 4.4 and 4.5~$\mu$Hz. The envelope of the beating should have a frequency of 0.1\,$\mu$Hz, i.e. 116\,days. By computing the Fourier transform of the magnetic proxy, we obtain a peak at 0.13\,$\mu$Hz, i.e. 89\,days. So the modulation observed is probably due to the beating of two frequencies. KIC~9226926 also has two main peaks in the power spectrum that have frequencies of 5.237\,$\mu$Hz and 5.259\,$\mu$Hz. The beating effect leads to a frequency of 0.022\,$\mu$Hz, i.e. 513\,days. We also computed the Fourier transform of the magnetic proxy and found a frequency of 0.025\,$\mu$Hz, i.e. 463 days. This also confirms that we are not observing the magnetic cycles of the star but a result of the beating effect. The detection of two frequencies for the rotation measurement implies that for these two stars we are observing the signature of the latitudinal differential rotation.

\begin{figure}[htbp]
\begin{center}
\includegraphics[angle=90, width=9cm]{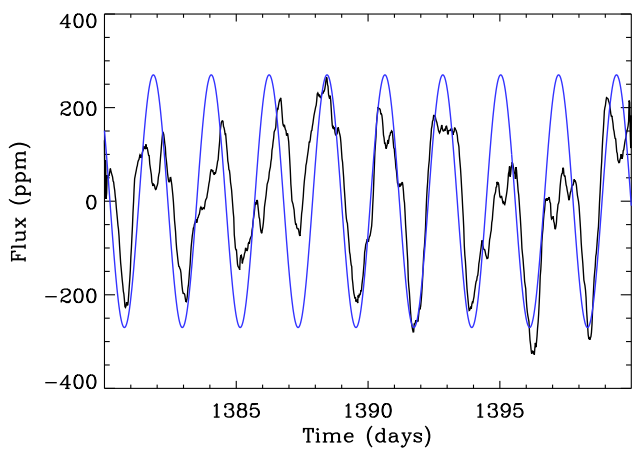}
\caption{Light curve of KIC 3733735 at a maximum of the amplitude
(black). The blue line is a sine wave whose frequency corresponds to
the maximum of the light curve spectrum. The truncated upper parts of
the real lightcurve are clearly visible while the lower parts are very
similar to the pure sine wave. This shape is characteristic of
starspots with surface rotation.}
\label{SK_spots}
\end{center}
\end{figure}

We also looked at the phase of these features and noticed that their phases are very stable with time, i.e. the features are present around the same region on the surface for a long period of time (approximately 1000 days). 
These long-lived features can be produced by long-lived
spots or active regions, by the presence of active longitudes yielding to the appearance of spots at a given longitude, or by classical pulsations. We favour the first two
explanations for two reasons. According to their fundamental parameters, we do not expect these stars to be in the instability strip. In addition, the morphology of
their light curves is typical of spotted stars. If we were in the presence of pulsations, the
modulation of the light curve would be strictly sinusoidal possibly with
a modulation due to starspots superimposed. However what we observe is a
nearly sinusoidal signal during the minima of the modulation and a
 sine wave more or less truncated at the maxima. This truncated
sinusoidal signal is characteristic of starspots, where the nearly flat part
corresponds to the continuum of the luminosity because of a low number
of spots. Figure~\ref{SK_spots} illustrates this with 20 days of data for KIC~3733735,
where this effect is clearly visible. This figure can be
directly compared to the one obtained through simulation by \citet{2013ApJS..205...17W}
(their Fig.1). They also point out that the visibility of this
continuum increases with the inclination of the star, a
correlation corroborated by the inclination angle of 30$^\circ$ for this star.

After ruling out the pulsation scenario, two explanations remain: the active regions are very stable or they are generated at the same longitude, also called the active longitudes. These active longitudes or active nests have been observed in the Sun \citep{2000ApJ...529.1101D} and seem to last for several rotation periods.  This has also been seen in stars with a nearby companion \citep[e.g.][]{2011A&A...533A..44L} whose interaction would generate a stable spot on the surface of the star. However, no signature of such a companion has been detected for these two stars. We notice that the long-term feature appears in KIC~3733735   once the second modulation (attributed to a magnetic cycle here) increases. This would mean that these active longitudes are related to internal changes, such as the dynamo mechanisms occurring inside the star. In particular this has been studied theoretically with dynamo models \citep{2006A&A...445..703B,2013ApJ...770..149W}.



\subsection{Magnetic index and other parameters}

Having defined a magnetic index and a magnetic cycle proxy for all the stars, we can study if the combination of both pieces of information can help to better understand the magnetic activity of the stars and the existence of regular cycles. 

The two stars that show an increase or decrease in activity, i.e. a slope in the magnetic proxy, have a magnetic index $ \langle S_{\rm ph} \rangle$ lower than the maximum activity of the Sun ($<S_{\rm ph}(n_{\rm high})>$) so unfortunately we cannot say anything more about these stars. We also notice that stars with high amplitude peaks in the magnetic proxy usually have larger magnetic indexes, while the stars with a more erratic behaviour have lower indexes in general (although with a few outliers). 

Finally, the two stars with broad peaks in the magnetic proxy (KIC~7668623 and KIC~10644253) have the highest values of $ \langle S_{\rm ph} \rangle$. As said in Section 4.2, these stars might reflect a configuration where the inclination angle of their rotation axis is low but high enough so that we can see the signature of the rotation. 

For 16 stars in our sample, \citet{2012MNRAS.423..122B} could estimate a v$\sin i$. Using these values in combination with the estimation of the radius of the star from scaling relations and the rotation periods we have inferred, we can deduce the inclination angle $i$. 
It appears that in general $i$ varies between 20 and 60$^{\circ}$. For three stars (KIC~7940546, 10016239, and 12009504), the $\sin i$ was a little larger than 1 but also slightly lower than 1 with the uncertainties. So we assumed that their inclination angle must be close to 90\,$^\circ$. This could be due to some errors in the seismic parameters. The angle of inclination is an important parameter because it affects the observation of the magnetic cycle as studied by \citet{2011JPhCS.271a2056V}. They showed that the maximum occurs later and has a lower amplitude than in the 90\,$^\circ$ case. This means that the minimum appears to be longer. In addition, they noted that the second harmonic of the rotation disappears. The star KIC~4638884 has an inclination angle around 25$^\circ$ and this could explain the shape of the magnetic proxy. The other star with a similar proxy (KIC~10644253) has an inclination angle of $\sim$\,43\,$^\circ$. The large uncertainty associated with this angle does not rule out the low inclination angle scenario. 

One thing that is unknown for the stars is the latitude where the starspots or active regions appear. If the star is observed pole on and the active regions appear close to the pole, then we would be sensitive to the lifetime of the active region. If the active latitudes are located around 30$^\circ$ like the Sun and the star is observed almost pole on, we would still be able to determine the surface rotation period of the star but the effect of the active region will be weakened in the light curve because of projection effects that will diminish the observed size of the pattern.  

\begin{figure}[htbp]
\begin{center}
\includegraphics[angle=90, width=9cm]{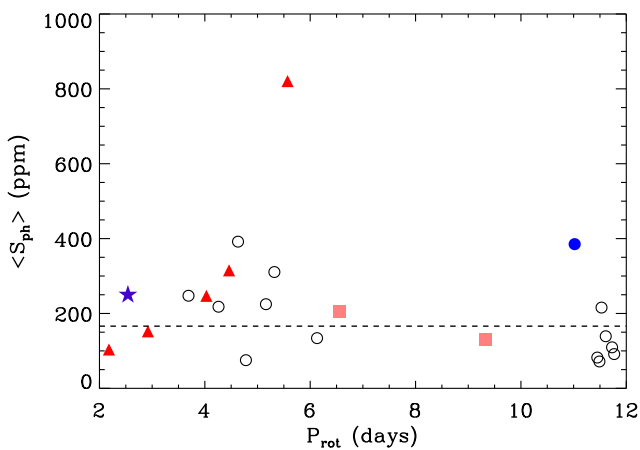}
\caption{Magnetic index $ \langle S_{\rm ph} \rangle$ versus $P_{\rm rot}$ using the same legend as in Figure 3. The dashed line represents the magnetic index for the Sun.}
\label{Sph_Prot}
\end{center}
\end{figure}

 A relation between the S-index and the rotation period has been established by \citet{2002AN....323..357S}. In Figure~\ref{Sph_Prot} we show $ \langle S_{\rm ph} \rangle$ as a function of $P_{\rm rot}$. At first sight no clear correlation appears, but if we look at the stars that show evidence of long-lived features (red triangles), a slight correlation seems to be present.  \citet{2002AN....323..357S} observed the opposite, that there was an anti-correlation, but we note that their result concerns stars where they observed magnetic cycles while our subsample only shows long-lived features.

We looked for evidence of correlation between the  magnetic index of our stars ($ \langle S_{\rm ph} \rangle$) and a few stellar parameters we have access to. It would be interesting to see if the hottest stars have generally larger magnetic indexes or not. According to \citet{2011arXiv1107.5325L}, who measured the CaHK index $R'_{\rm HK}$ for 300 FGK, stars with an effective temperature above 5600K showed a magnetic cycle semi-amplitude smaller than for cooler stars. The F stars in our sample show a slight anti-correlation where hotter stars have smaller magnetic indexes, going in the same direction as the \citet{2011arXiv1107.5325L} results. 


We expect that magnetic activity declines with age. The mean large separation $\Delta \nu$ gives a hint about the evolutionary stage of a star. However, we did not find any clear correlation between $ \langle S_{\rm ph} \rangle$ and $\Delta \nu$.

\begin{figure}[htbp]
\begin{center}
\includegraphics[angle=90, width=9cm]{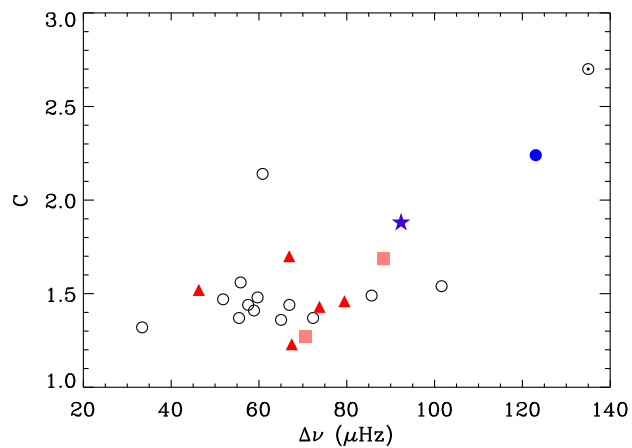}
\caption{Contrast between maximum and minimum activity regions as a function of the mean large separation ($\Delta \nu$). Using the same legend as in Figure 3. The Sun is represented with the symbol $\odot$.}
\label{C_Dnu}
\end{center}
\end{figure}

The contrast as a function of the evolutionary stage (Fig.~\ref{C_Dnu}) presents a slight correlation, which could suggest that more evolved stars seem to show a smaller contrast, which means that the change in luminosity due to magnetic activity becomes less prominent. However, the correlation rate is 40\%, which is not negligible. We still have to be cautious because for most of the stars we do not see a complete cycle, if such a cycle exists in all F stars, which could bias our analysis.


\subsection{Rossby number}

\citet{1984ApJ...279..763N} studied the magnetic activity of low main-sequence stars that had \ion{Ca}{ii} H and K line measurements at the Mount Wilson. They showed that chromospheric activity directly depends on the Rossby number Ro= $P_{\rm rot}/\tau_c$, where $\tau_c$ is the convective turnover time. \citet{2002AN....323..357S} showed that the magnetic activity cycle period and the rotation rate were also linked through this quantity.
We computed the Rossby numbers for our sample of stars. To ease comparisons with previous works, we computed a convective turnover time from the colour $B-V$, as done in \citet{1984ApJ...279..763N}. Colors are derived from the Tycho2 catalog \citep{2000A&A...355L..27H}. We have to keep in mind, however, that this approach, where $\tau_c$ is only parametrized with a color, is very crude. Moreover, it is not necessarily well suited for F stars that may have very shallow convective envelopes. These numbers are quoted in Table~\ref{tbl-3}.

However, seismology allows us to perform accurate modelling of the internal structure of observed stars. It is possible to compute Rossby numbers from these tailored models. Eight stars have been modelled by the Asteroseismic Modeling Portal \citep[][Metcalfe et al. in prep.]{2009ApJ...699..373M,2012ApJ...749..152M}. Using the mixing length theory, we defined a local Rossby number $Ro_{l}$ and a global Rossby number $Ro_{g}$. For the first, the convective turnover time scale is computed as $\tau_{cl} = \Lambda / v_{\mathrm{MLT}}$, where $\Lambda$ is the mixing length and $v_{\mathrm{MLT}}$ is the convective velocity evaluated half a pressure scale height above the base of the convective envelope. For $Ro_{g}$, the convective turnover time scale is computed through the whole envelope as $\tau_{cg} = \int_{r_{\mathrm{bce}}}^{R} dr/v_{\mathrm{MLT}}$, where $r_{\mathrm{bce}}$ is the position of the base of the convective envelope and $R$ the stellar radius. These results are given in Table~\ref{tbl-3}. We verify that $Ro_{l}$ and $Ro_{g}$ are closely linked and vary similarly. We note that the large values for the Rossby number of 12 and 125 correspond to the stars that have the shallowest convective zones (depths of 5 and 2\%).

Among the two stars for which a cycle has been observed, only KIC~3733735 has seismic constraints that allow structure modelling. The Rossby number found for this star is high enough to suggest a long cycle period $P_{\mathrm{cyc}}$ compared to its rotation period $P_\mathrm{rot}$. Indeed, by renormalizing the cycle period to the rotation period for this star we find $P_{\mathrm{cyc}} > 540 P_\mathrm{rot}$ (since only one cycle was observed), whereas for the Sun, $P_{\mathrm{cyc}} \approx 150 P_\mathrm{rot}$. These observations should give constraints for dynamo theory that makes predictions for the link between $P_{\mathrm{cyc}}/P_{\mathrm{rot}}$ and $Ro$ \citep[see][]{2008sust.book.....T}.
 
\begin{figure}[htbp]
\begin{center}
\includegraphics[width=6.5cm, angle=90]{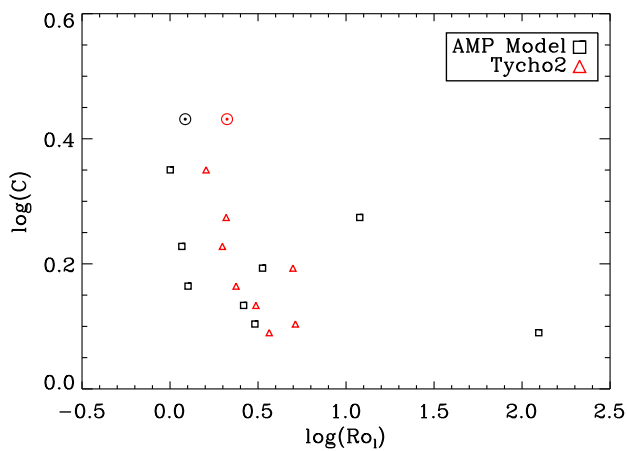}
\caption{Magnetic contrast C as a function of the Rossby number of the stars modelled by AMP. The Rossby number was obtained from the stellar models (squares) and with the equation of Noyes et al. (1984) based on colour measurements of Tycho2 (triangles). }
\label{Rossby}
\end{center}
\end{figure}

We finally searched for a correlation between the magnetic index and the local Rossby number. \citet{1984ApJ...279..763N} showed that there was a clear anti-correlation between the Ca II HK flux $R'_{HK}$ and Ro. Recently,  \citet{2013arXiv1311.3374M} analysed a large number of solar-type stars with high-resolution spectropolarimetric observations and found a similar anti-correlation between the strength of the magnetic field and Ro. However, with our sample of stars we could not find any significant anti-correlation between $ \langle S_{\rm ph} \rangle$, and the local Rossby number. However, we found a small anti-correlation between the contrast C and the Rossby number (see Figure~\ref{Rossby}); this result does not agree with the results of \citet{2002AN....323..357S} who did not find any clear correlation between the cycle amplitude and Ro.


\subsection{Hints in the background}

It has been observed that the increase in the Sun's magnetic activity
is usually accompanied by the presence of bright points (or
faculae) \citep[e.g.][]{2012MNRAS.421.3170K}. To investigate the background of the
power spectrum and extract its parameters for the 22 stars,
we fitted it following the A2Z method described in \citet{2011ApJ...741..119M}.

In order to have the most complete power spectrum, we used
the short cadence data allowing us to measure the photon noise.
Following \citet{2013ApJ...767...34K} we fitted the background with one and two Harvey-like  functions \citep{1985ESASP.235..199H}
by fixing the slopes to 3.5 for the granulation component and
6.2 for the higher frequency component (i.e. the bright points) to
ensure the stability of the fits. According to \citet{2012MNRAS.421.3170K}, the second Harvey component could be due to faculae, then linked to the magnetic activity. However, within our sample of F stars, there is no evidence of a correlation between the magnetic index $ \langle S_{\rm ph} \rangle$ and
the characteristic parameters of the second Harvey component.

\begin{table*}[htdp]
\caption{Table with rotation periods, inclination angle, and Rossby numbers for the 22 stars computed with different approaches: $R_{ol}$ and  $R_{og}$ are local and global Rossby numbers computed from tailored models, while $R_o$(N84) is the Rossby number derived from the $B-V$ color as in \citet{1984ApJ...279..763N}. The last column contains the depth of the convective zone (as a function of the stellar radius) when available with asteroseismology.}
\begin{center}
\begin{tabular}{rrrccccccc}
\hline
\hline

  Ref. no & KIC& $P_{\rm rot}$ (days) & $i$ ($^\circ$) & $R_{ol}$  & $R_{og}$ & $R_o$ (N84) & $d_{\rm CZ}$ (R)\\
\hline
1&  1430163 &   3.88\,$\pm$\, 0.58 &  31.75\,$\pm$\, 7.57&         &        &  1.54  \\
2&  1435467 &   6.53\,$\pm$\, 0.64 &  47.08\,$\pm$\,10.92&   3.026 &  2.248 &  5.14  & 0.12\\
3&  3733735 &   2.54\,$\pm$\, 0.16 &  31.41\,$\pm$\, 3.52&  11.964 &  8.762 &  2.08 & 0.05 \\
 4& 4638884 &   6.17\,$\pm$\, 0.60 &  27.23\,$\pm$\, 5.27&         &        &  2.65  \\
 5& 5371516 &   5.13\,$\pm$\, 0.48 &  47.63\,$\pm$\, 9.21&         &        &  1.68  \\
 6& 5773345 &  11.60\,$\pm$\, 0.90 &  47.47\,$\pm$\,11.97&         &        &  4.24  \\
 7& 6508366 &   3.78\,$\pm$\, 0.34 &  36.92\,$\pm$\, 5.50&         &        &  2.20  \\
 8& 7103006 &   4.61\,$\pm$\, 0.38 &  35.70\,$\pm$\, 5.46&         &        &  1.21  \\
 9& 7206837 &   4.04\,$\pm$\, 0.25 &  29.76\,$\pm$\, 4.06&   1.269 &  1.036 &  2.38  & 0.13\\
 10& 7668623 &   5.12\,$\pm$\, 0.38 &  25.13\,$\pm$\, 3.73&         &        &  0.99  \\
 11& 7771282 &  11.91\,$\pm$\, 0.96 &  ...                &         &        & 40.93  \\
 12& 7940546 &  11.12\,$\pm$\, 0.85 &   90\,$\pm$\,20      &         &        &  4.21  \\
 13& 8026226 &  11.14\,$\pm$\, 2.16 &  47.57\,$\pm$\,19.80&         &        &  4.17  \\
 14& 9226926 &   2.17\,$\pm$\, 0.14 &  50.31\,$\pm$\, 6.20&         &        &  2.59  \\
 15& 9289275 &  11.31\,$\pm$\, 1.02 &   ...               &         &        &  2.99  \\
 16& 9812850 &   5.05\,$\pm$\, 0.53 &  44.33\,$\pm$\, 9.24&   2.563 &  2.047 &  3.01 & 0.12 \\
17& 10016239 &   4.92\,$\pm$\, 0.38 &  90\,$\pm$\,12       &         &        &  1.13  \\
18 & 10162436 &  11.59\,$\pm$\, 0.88 &  47.06\,$\pm$\,11.72&   3.303 &  2.267 &  4.91  & 0.15\\
19& 10355856 &   4.47\,$\pm$\, 0.28 &  20.39\,$\pm$\, 3.14& 124.707 & 70.213 &  3.67  & 0.02\\
20& 10644253 &  10.89\,$\pm$\, 0.81 &  43.44\,$\pm$\,14.48&   0.991 &  0.717 &  1.59  & 0.22\\
21& 11070918 &   2.91\,$\pm$\, 0.18 &  ...                &         &        &  2.39  \\
22& 12009504 &   9.48\,$\pm$\, 0.85 &   90\,$\pm$\,27      &   1.185 &  0.903 &  2.02  & 0.20\\
 & Sun & 26 & 90 & 1.222 & 0.923 & 2.108& 0.30\\
\hline

\hline
\end{tabular}
\end{center}
\label{tbl-3}
\end{table*}%


\section{Conclusions}

We have analysed 22 F stars that have a rotation period shorter than 12 days to determine their magnetic indexes and study their magnetic proxies. We defined some new indicators of the magnetic activity based on photometric observations. We studied the magnetic index $\langle S_{\rm ph} \rangle$, which is the mean value of the standard deviations of the subseries of length $5 \times P_{\rm rot}$. We showed that in general, most of the stars we studied have a mean activity index that is slightly larger than the one of the Sun. 

We combined the measurement of the magnetic index with the magnetic proxy obtained with the wavelet analysis. This analysis showed a variety of behaviour in the magnetic activity of the F stars. Two stars show cycle-like behaviour. The magnetic proxy of KIC~3733735  is clearly flat over a period of 500 days and shows an increase and decrease during the rest of the observations suggesting a cycle period of at least 1400 days. The other star, KIC~10644253, has a magnetic proxy with two large peaks that can be explained by the observation of two cycles for a star that has a low inclination angle.

For two stars, KIC~3733735 and KIC~9226296, we detected the signature of latitudinal differential rotation as we observed the beating of  two close frequencies. A group of five stars show a clear signature of long-lived features on their surface suggesting the existence of active longitudes on these stars.


We observed stars that present a slight increase or decrease in the magnetic proxy, suggesting a modulation of the magnetic activity. Finally, the last category gathers stars where no clear modulation is seen. These stars have either a large number of high amplitude peaks, or result from the observation with an inclination angle and more erratic stars. 

This is very interesting because it means that stars with similar spectroscopic constraints and rotation periods can have very different magnetic activity behaviour: cycle-like behaviour, long-lived features, chaotic activity, or flat activity. To understand these differences, it is important to measure the magnetic field of these stars but also have a detailed description of their structure, such as the depth of their convection zones. For the former point, four in our sample of stars (KIC~1435467, 3733735, 7206837, and 12009504) have been monitored by the FIES spectrograph at the Nordic Optical Telescope \citep{2013MNRAS.433.3227K}. The analysis agrees that the star with the largest average S-index of 0.182\,$\pm$\,0.001 is KIC~3733735. For the study of the structure a deeper analysis will be done for the stars with long short-cadence data.

\bibliographystyle{aa} 
\bibliography{/Users/Savita/Documents/BIBLIO_sav}

\begin{acknowledgements}
This work was partially supported by the NASA grant NNX12AE17G. SM and RAG acknowledge the support of the European Community's Seventh Framework Programme (FP7/2007-2013) under grant agreement no. 269194 (IRSES/ASK). RAG acknowledges the support of the French ANR/IDEE grant. The authors thank K. Augustson, A.~S. Brun, S. Mathis, and B. Mosser for very useful discussions. The research leading to these results has received funding from the European Research Council under the European Community's Seventh Framework Programme (FP7/2007--2013)/ERC grant agreement n$^\circ$227224 (PROSPERITY), as well as from the Research Council of the KU Leuven under grant agreement GOA/2013/012. SB is supported by the Foundation for Fundamental Research on Matter (FOM),
which is part of the Netherlands Organisation for Scientific Research (NWO).
 \end{acknowledgements}

\appendix

\section{Appendix}

\begin{figure}[htbp]
\begin{center}
\includegraphics[width=10cm, trim=0.5cm 6cm 4cm 0.5cm]{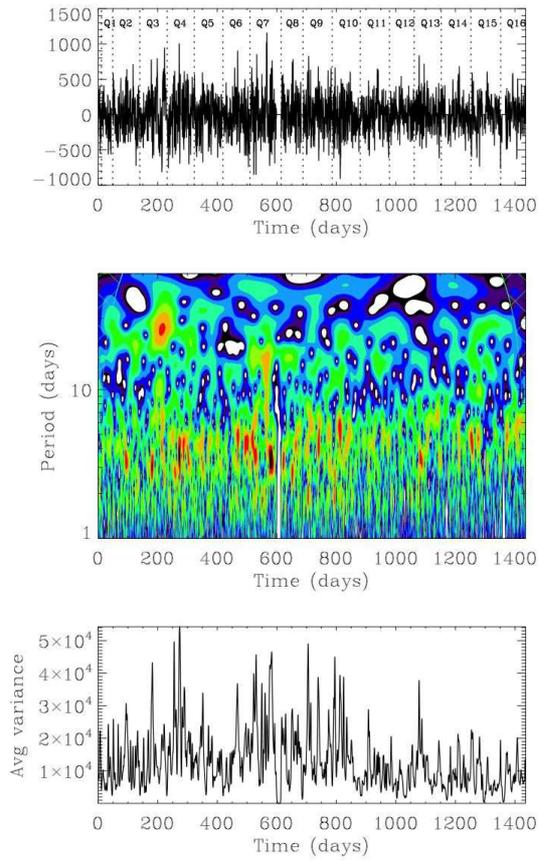}
\caption{ Wavelet analysis of KIC~1430163 {\bf (group O)}. Top panel: temporal variation of the flux after the corrections applied as in \citet{2011MNRAS.414L...6G}. Middle panel: Wavelet power spectrum as a function of time and period. The green grid represents the cone of influence that takes into account the edge effects and delimits the reliable rein in the WPS. Red and dark colours correspond to strong power while blue corresponds to weak power. Bottom panel: projection of the WPS on the time axis between periods of 1 and 7 days.}
\label{wavelets1}
\end{center}
\end{figure}

\begin{figure}[htbp]
\begin{center}
\includegraphics[width=10cm, trim=0.5cm 6cm 4cm 0.5cm]{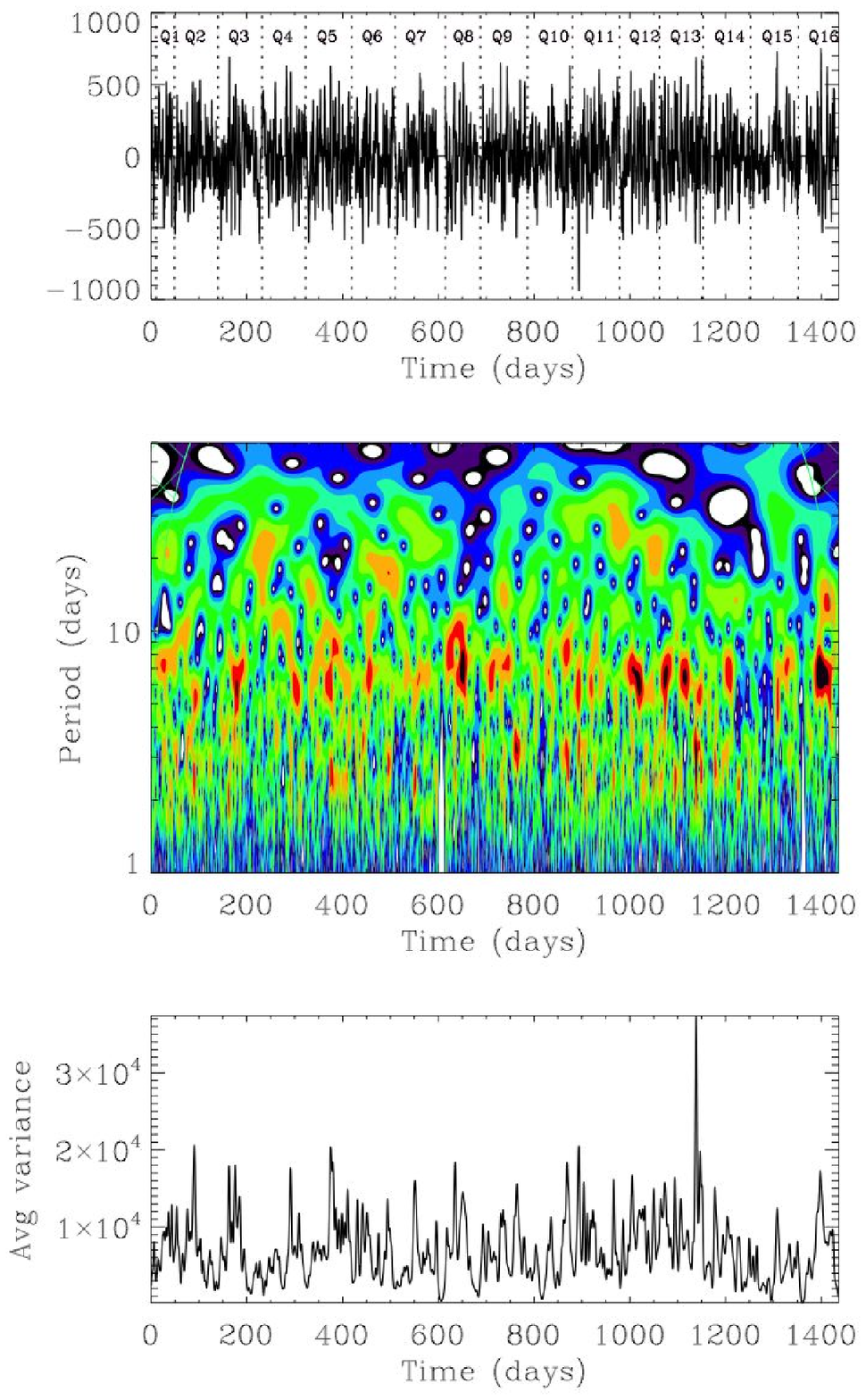}
\caption{ Wavelet analysis of KIC~1435467 {\bf (group T)}. Same representation as in Figure~\ref{wavelets1}. The magnetic proxy (bottom panel) was computed between periods of 1.7 and 11 days.}
\label{wavelets2}
\end{center}
\end{figure}

\begin{figure}[htbp]
\begin{center}
\includegraphics[width=10cm, trim=0.5cm 6cm 4cm 0.5cm]{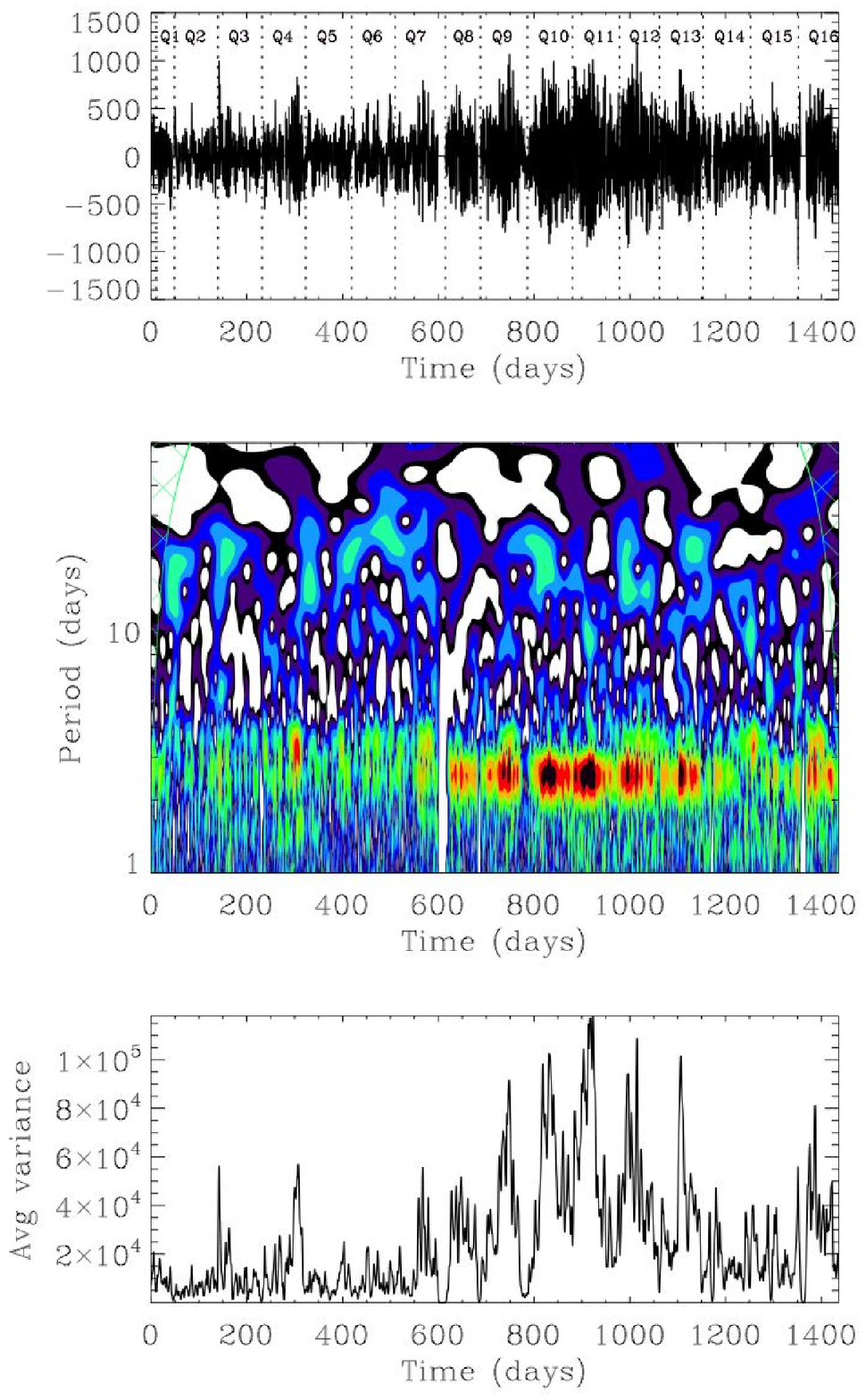}
\caption{ Wavelet analysis of KIC~3733735 {\bf (group L+C)}. Same representation as in Figure~\ref{wavelets1}. The magnetic proxy (bottom panel) was computed between periods of 1 and 4.1 days.}
\label{wavelets3}
\end{center}
\end{figure}

\begin{figure}[htbp]
\begin{center}
\includegraphics[width=10cm, trim=0.5cm 6cm 4cm 0.5cm]{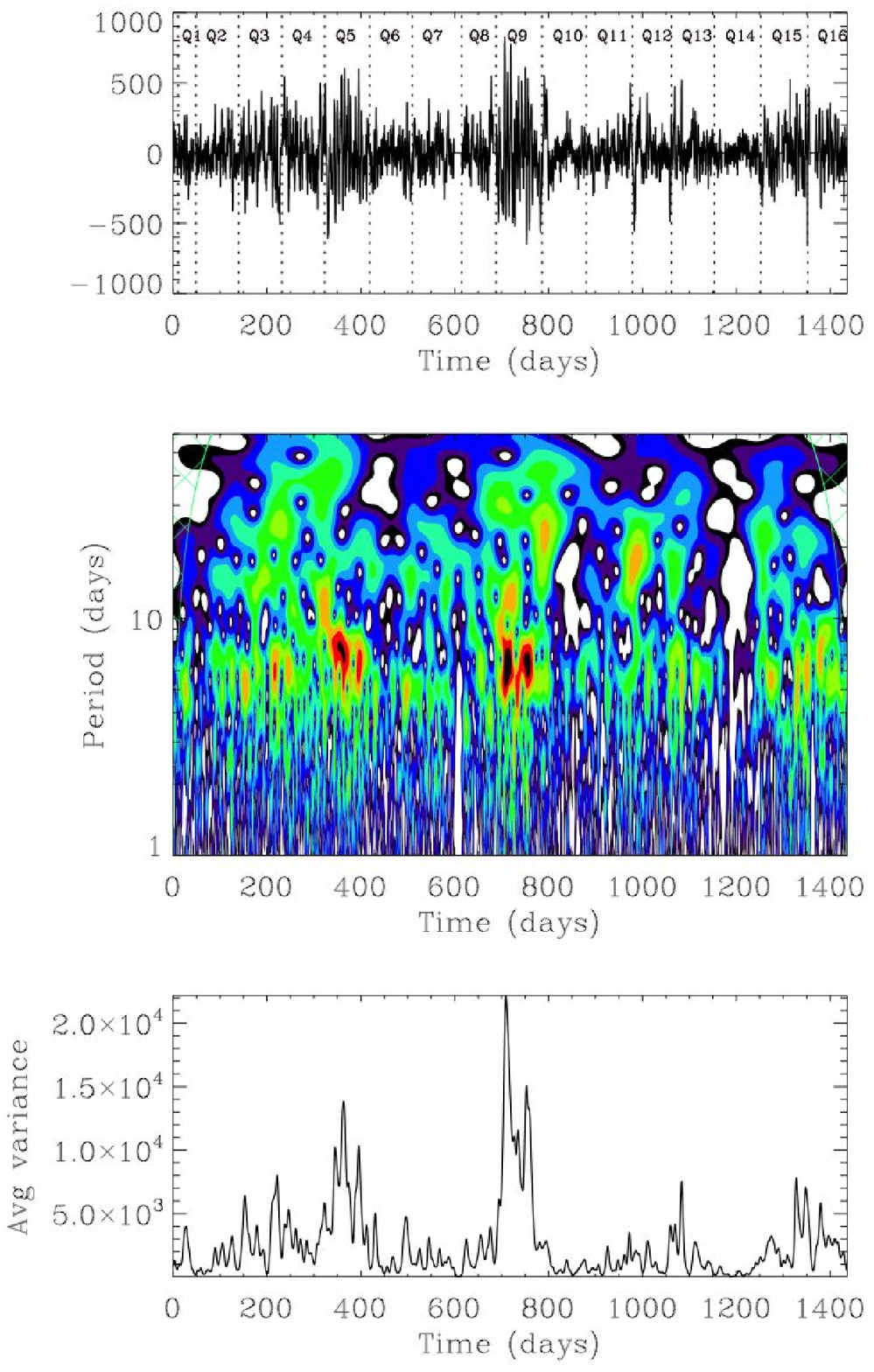}
\caption{ Wavelet analysis of KIC~4638884 {\bf (group O)}. Same representation as in Figure~\ref{wavelets1}. The magnetic proxy (bottom panel) was computed between periods of 2.5 and 12 days.}
\label{wavelets4}
\end{center}
\end{figure}

\begin{figure}[htbp]
\begin{center}
\includegraphics[width=10cm, trim=0.5cm 6cm 4cm 0.5cm]{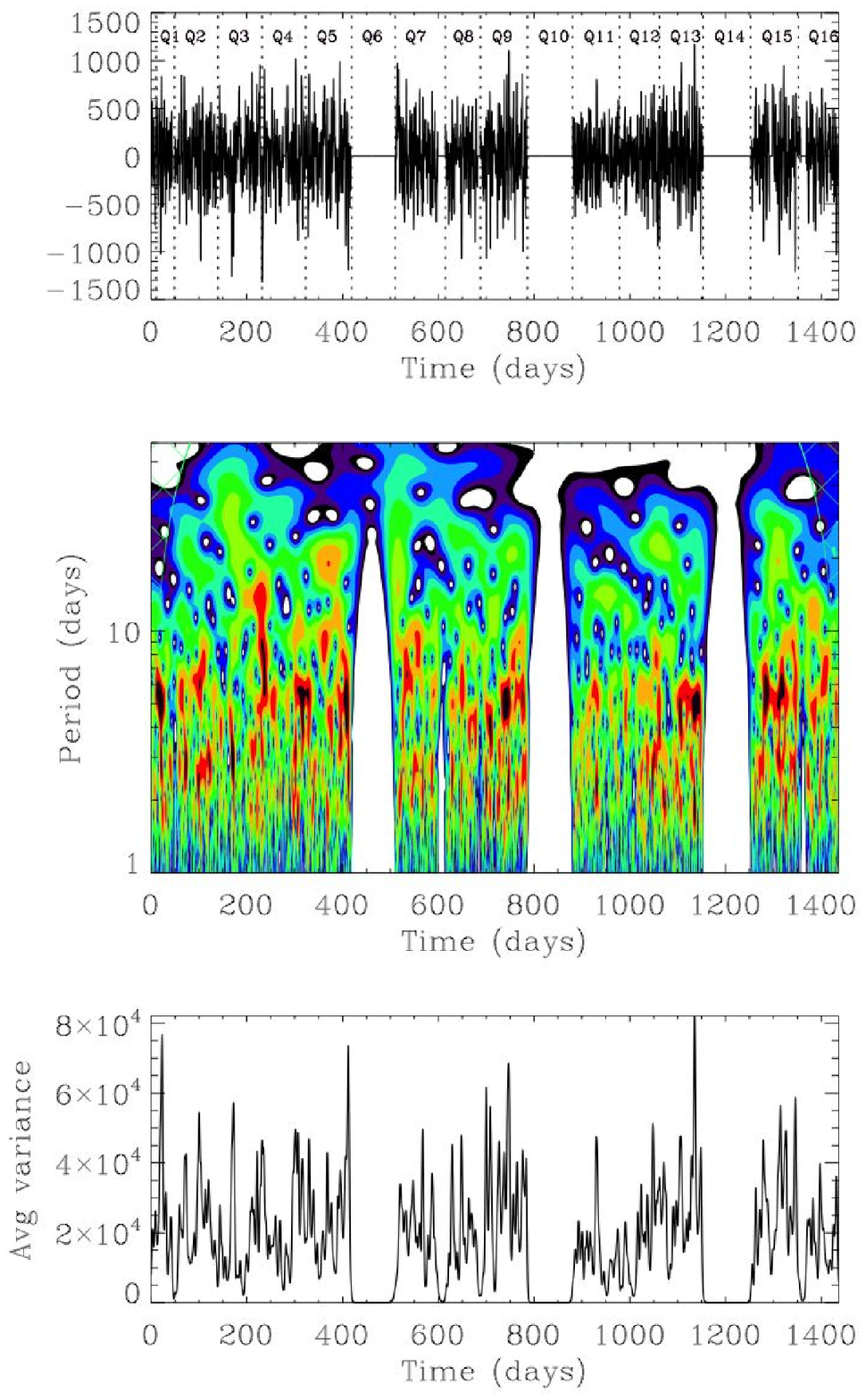}
\caption{ Wavelet analysis of KIC~5371516{ \bf (group O)}. Same representation as in Figure~\ref{wavelets1}. The magnetic proxy (bottom panel) was computed between periods of 1.5 and 10 days.}
\label{wavelets5}
\end{center}
\end{figure}

\begin{figure}[htbp]
\begin{center}
\includegraphics[width=10cm, trim=0.5cm 6cm 4cm 0.5cm]{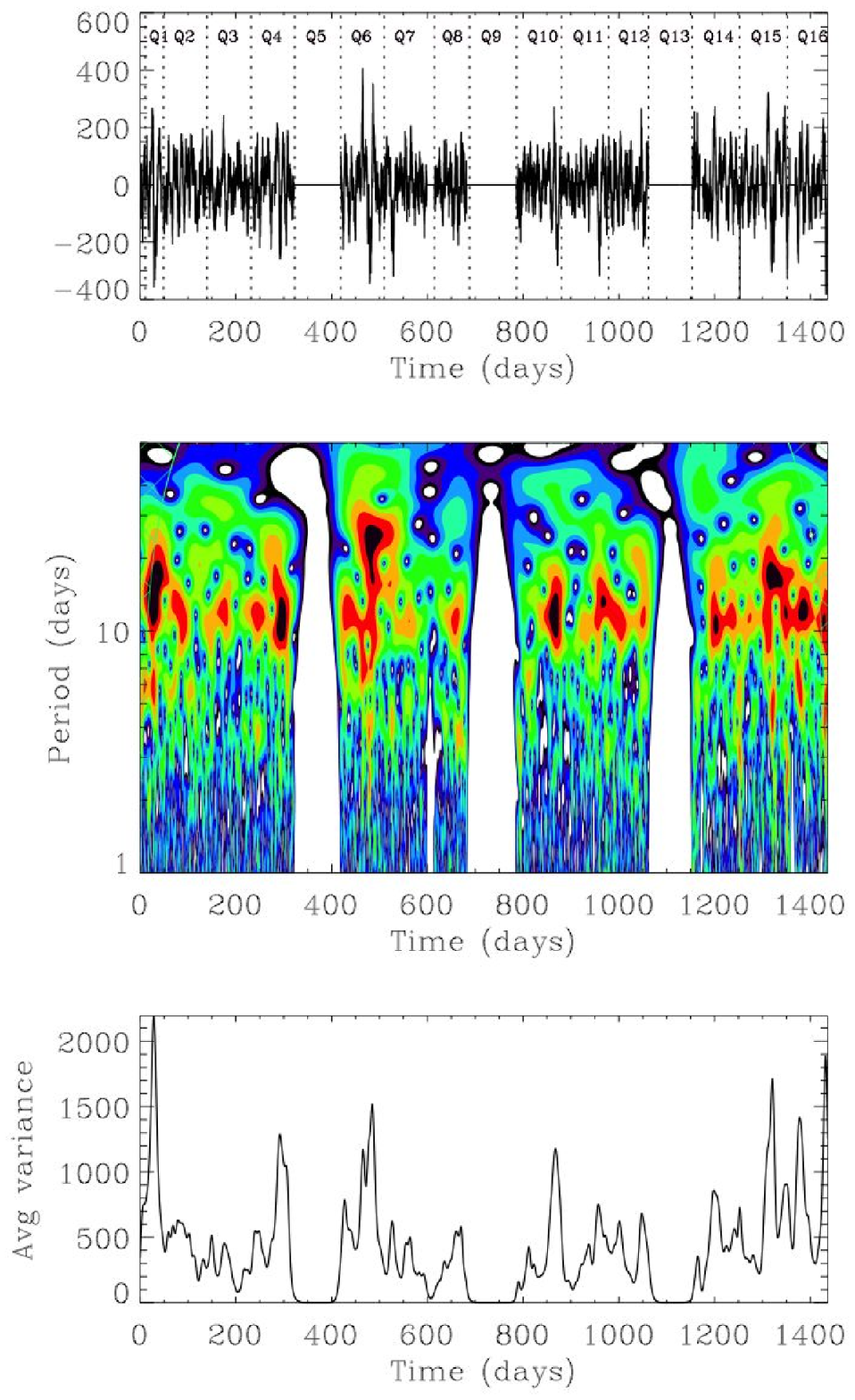}
\caption{ Wavelet analysis of KIC~5773345 {\bf (group O)}. Same representation as in Figure~\ref{wavelets1}. The magnetic proxy (bottom panel) was computed between periods of 3.8 and 20 days.}
\label{wavelets6}
\end{center}
\end{figure}

\begin{figure}[htbp]
\begin{center}
\includegraphics[width=10cm, trim=0.5cm 6cm 4cm 0.5cm]{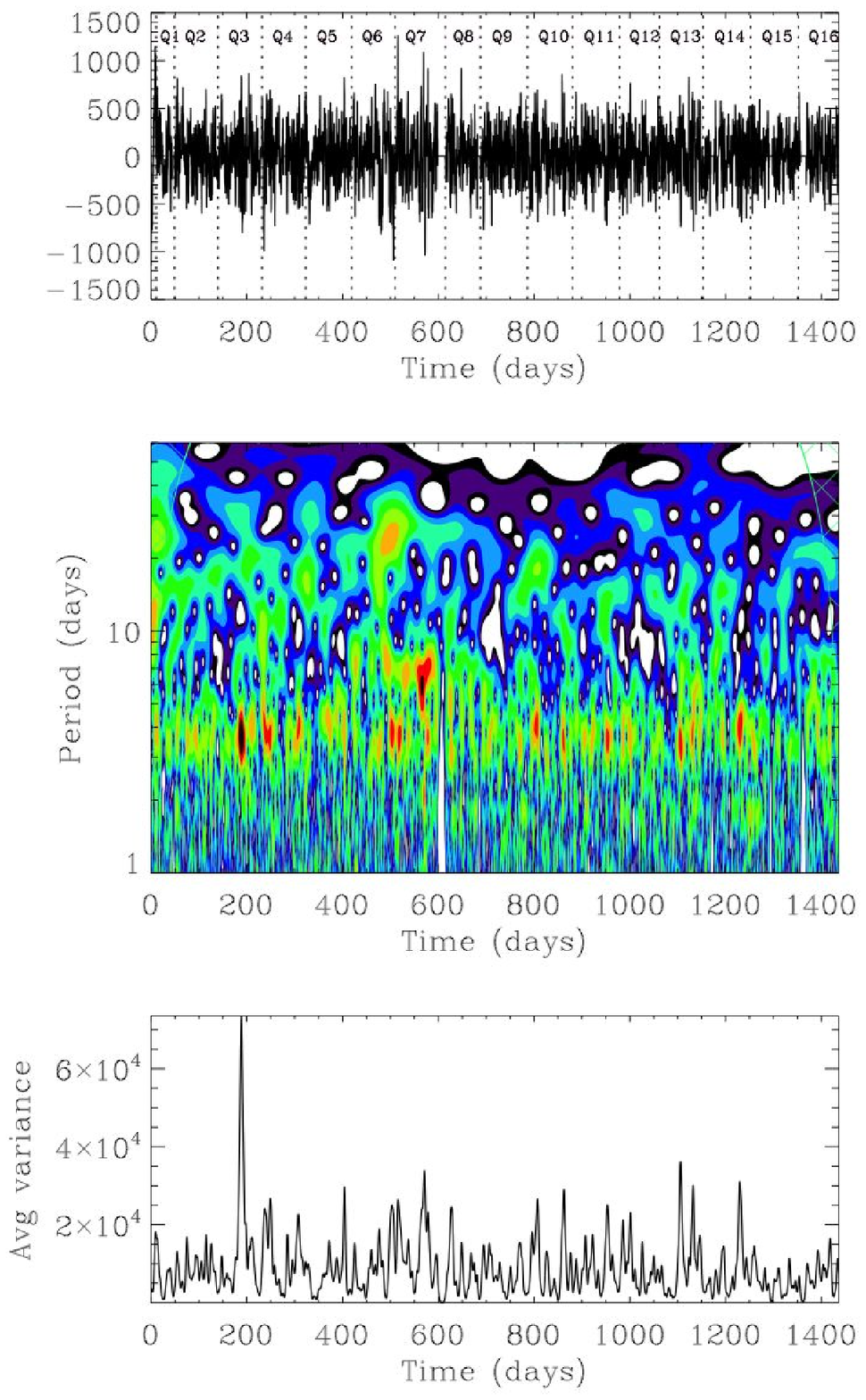}
\caption{ Wavelet analysis of KIC~6508366 {\bf (group O)}. Same representation as in Figure~\ref{wavelets1}. The magnetic proxy (bottom panel) was computed between periods of 2 and 6 days.}
\label{wavelets7}
\end{center}
\end{figure}

\begin{figure}[htbp]
\begin{center}
\includegraphics[width=10cm, trim=0.5cm 6cm 4cm 0.5cm]{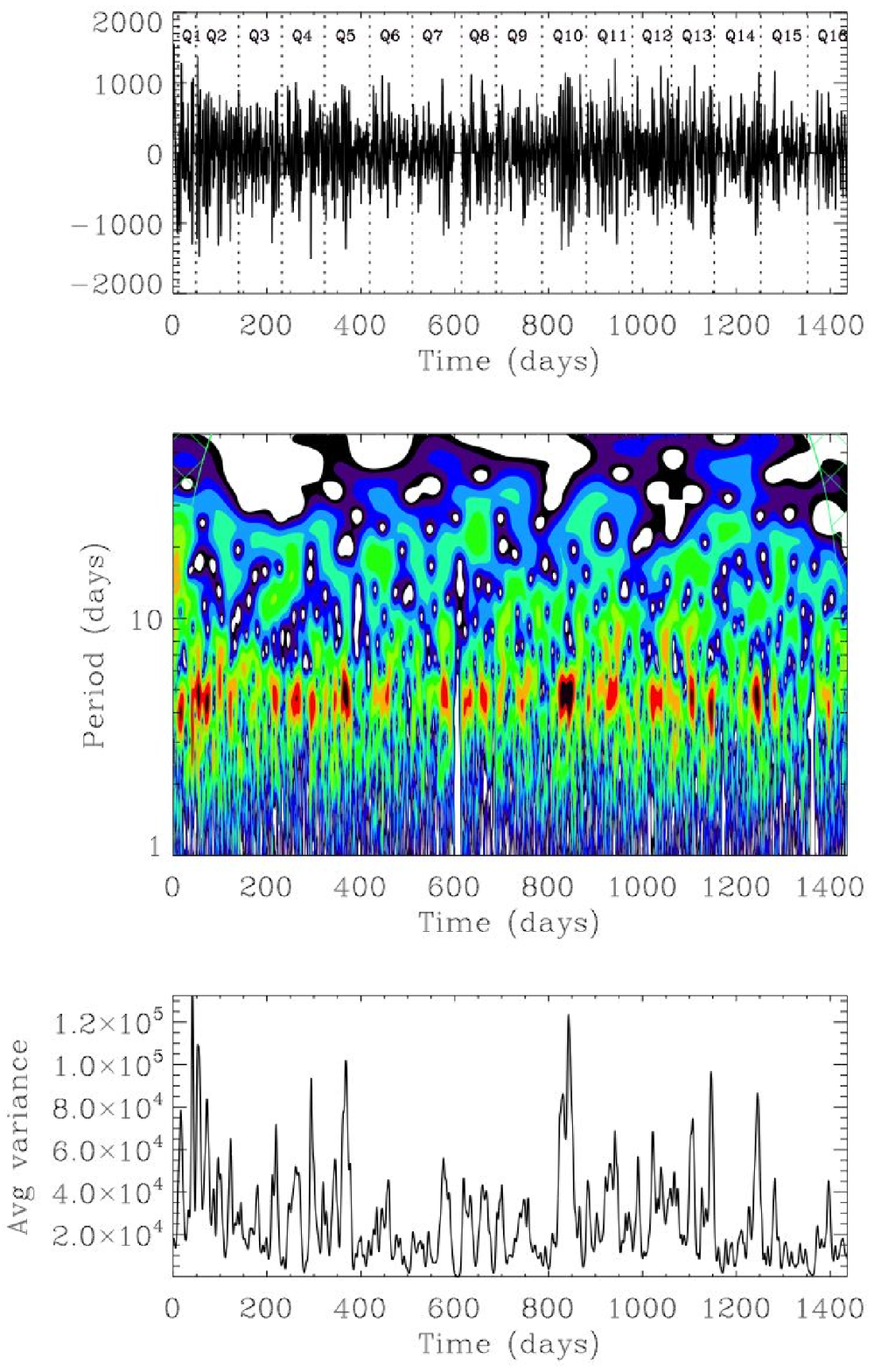}
\caption{ Wavelet analysis of KIC~7103006 {\bf (group O)}. Same representation as in Figure~\ref{wavelets1}. The magnetic proxy (bottom panel) was computed between periods of 2 and 8 days.}
\label{wavelets7}
\end{center}
\end{figure}

\begin{figure}[htbp]
\begin{center}
\includegraphics[width=10cm, trim=0.5cm 6cm 4cm 0.5cm]{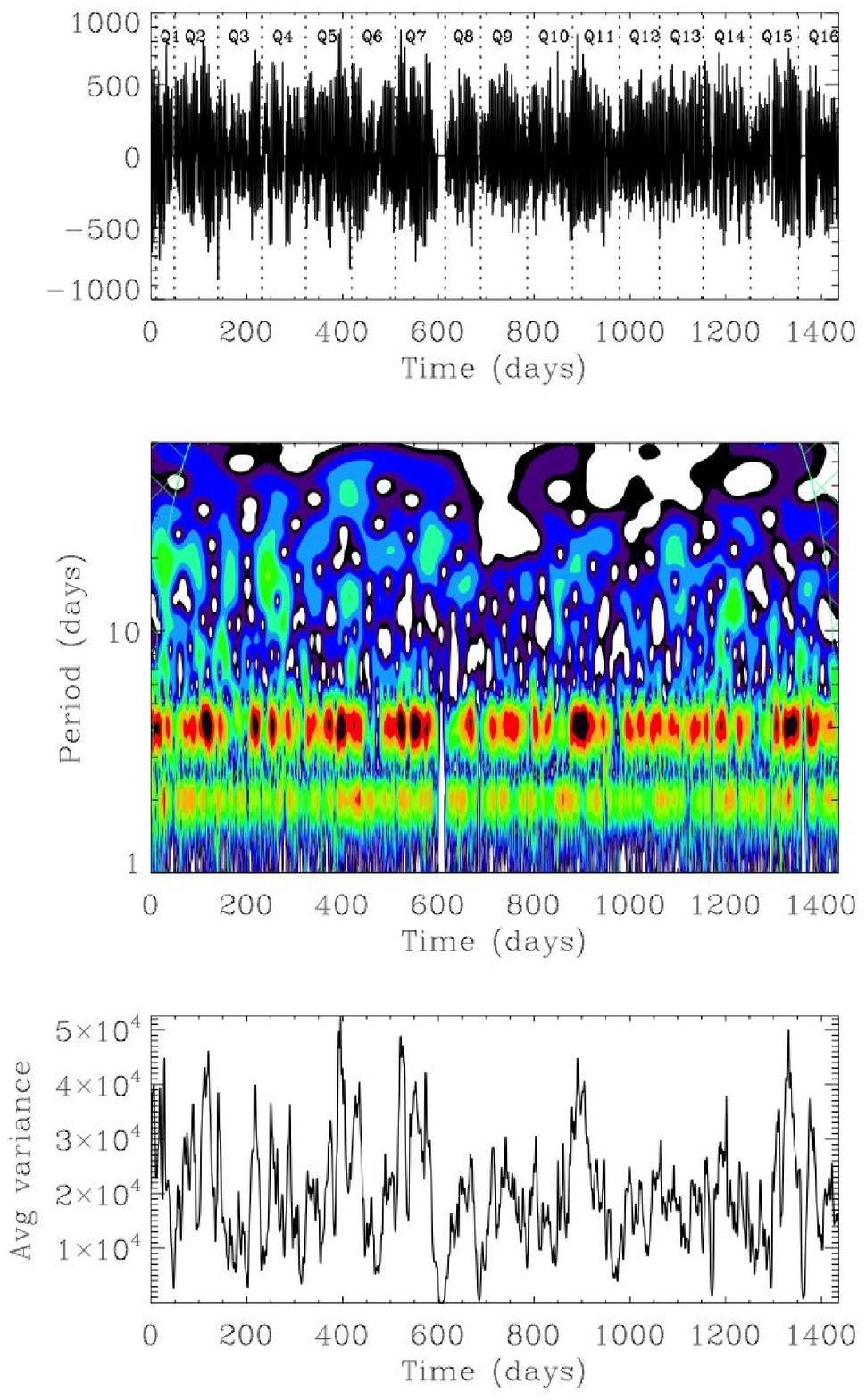}
\caption{ Wavelet analysis of KIC~7206837 {\bf (group L)}. Same representation as in Figure~\ref{wavelets1}. The magnetic proxy (bottom panel) was computed between periods of 1.2 and 7 days.}
\label{wavelets8}
\end{center}
\end{figure}

\begin{figure}[htbp]
\begin{center}
\includegraphics[width=10cm, trim=0.5cm 6cm 4cm 0.5cm]{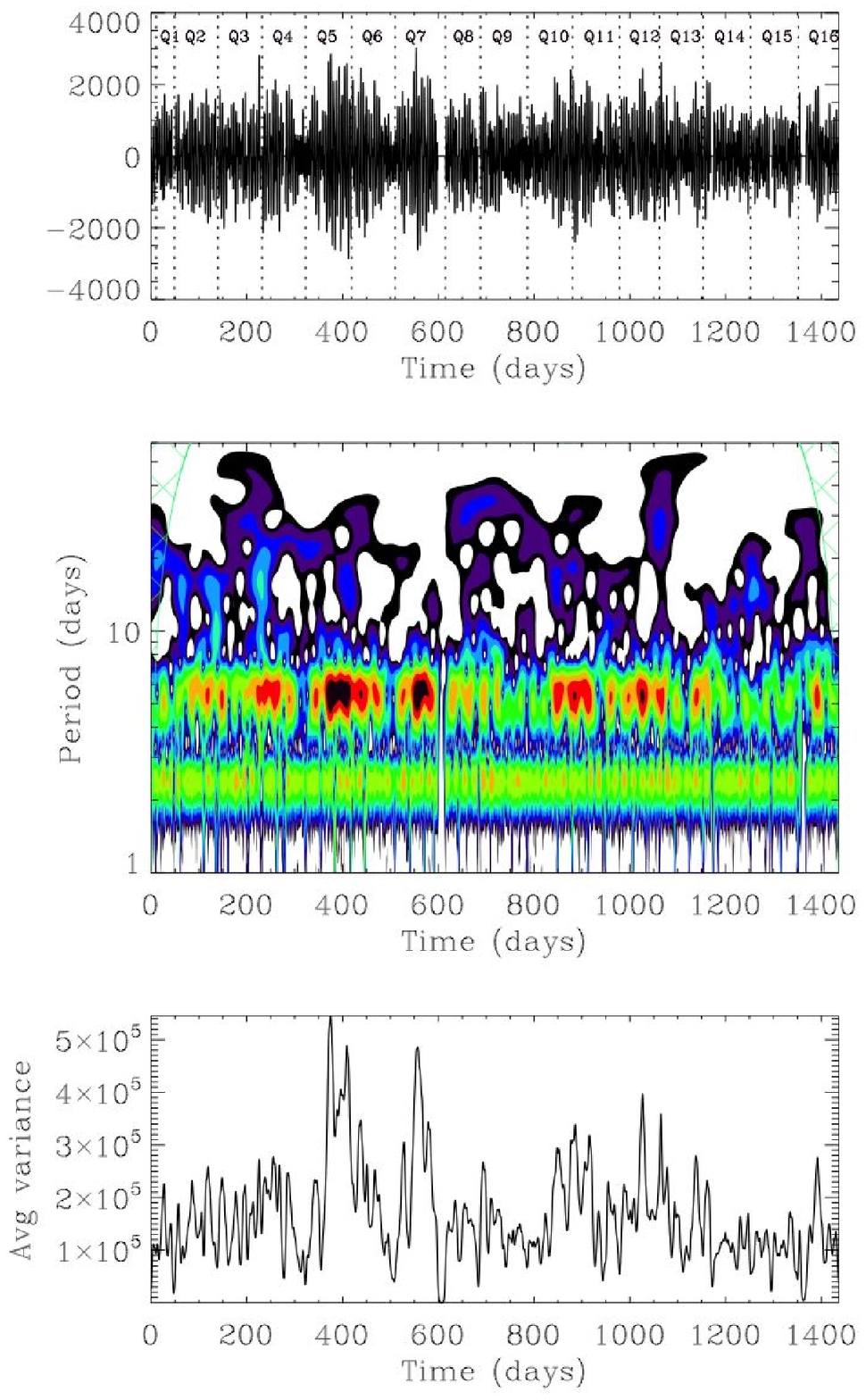}
\caption{ Wavelet analysis of KIC~7668623 {\bf (group L)}. Same representation as in Figure~\ref{wavelets1}. The magnetic proxy (bottom panel) was computed between periods of 1.9 and 8 days.}
\label{wavelets9}
\end{center}
\end{figure}

\clearpage

\begin{figure}[htbp]
\begin{center}
\includegraphics[width=10cm, trim=0.5cm 6cm 4cm 0.5cm]{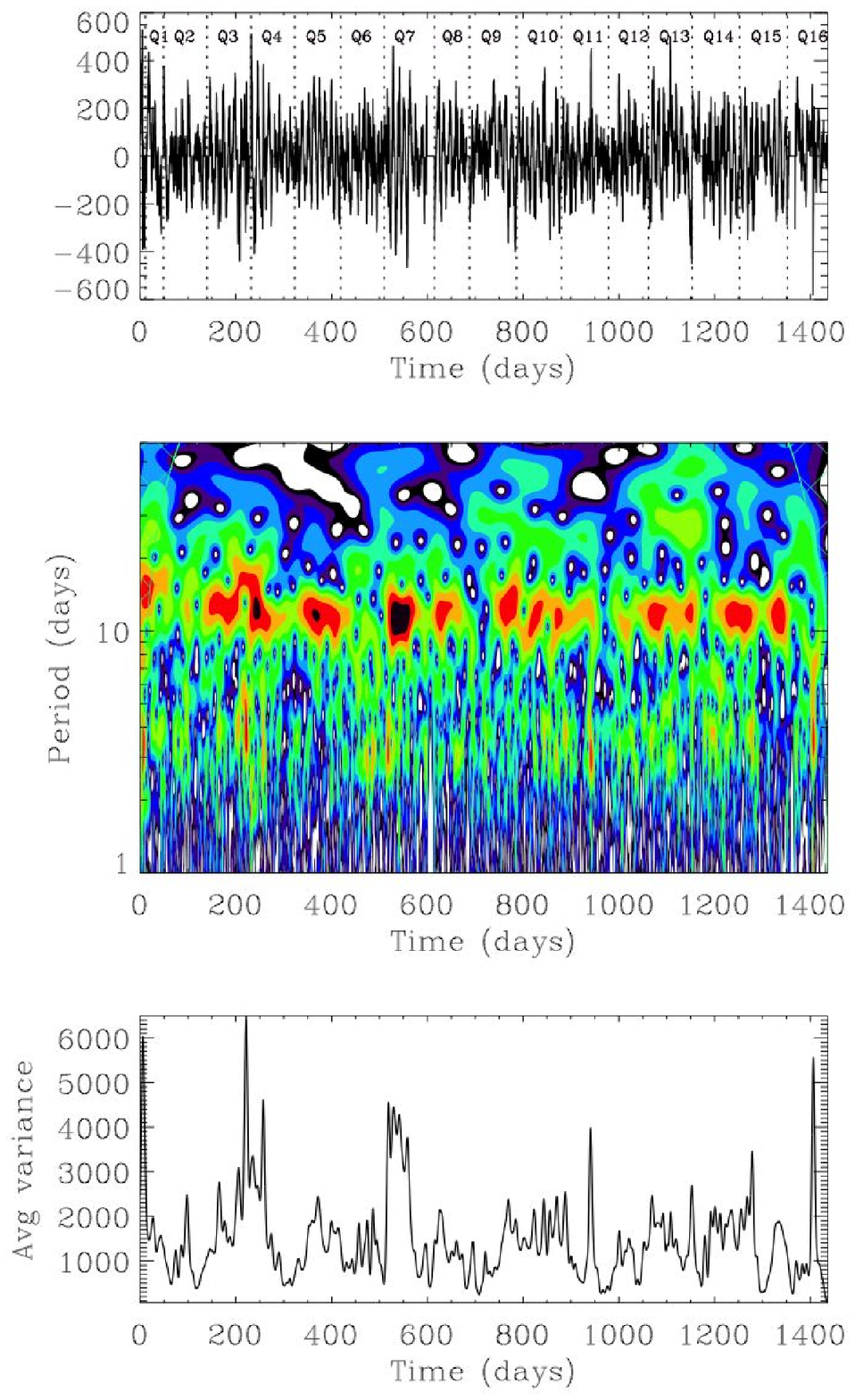}
\caption{ Wavelet analysis of KIC~7771282 {\bf (group O)}. Same representation as in Figure~\ref{wavelets1}. The magnetic proxy (bottom panel) was computed between periods of 3 and 20 days.}
\label{wavelets10}
\end{center}
\end{figure}

\begin{figure}[htbp]
\begin{center}
\includegraphics[width=10cm, trim=0.5cm 6cm 4cm 0.5cm]{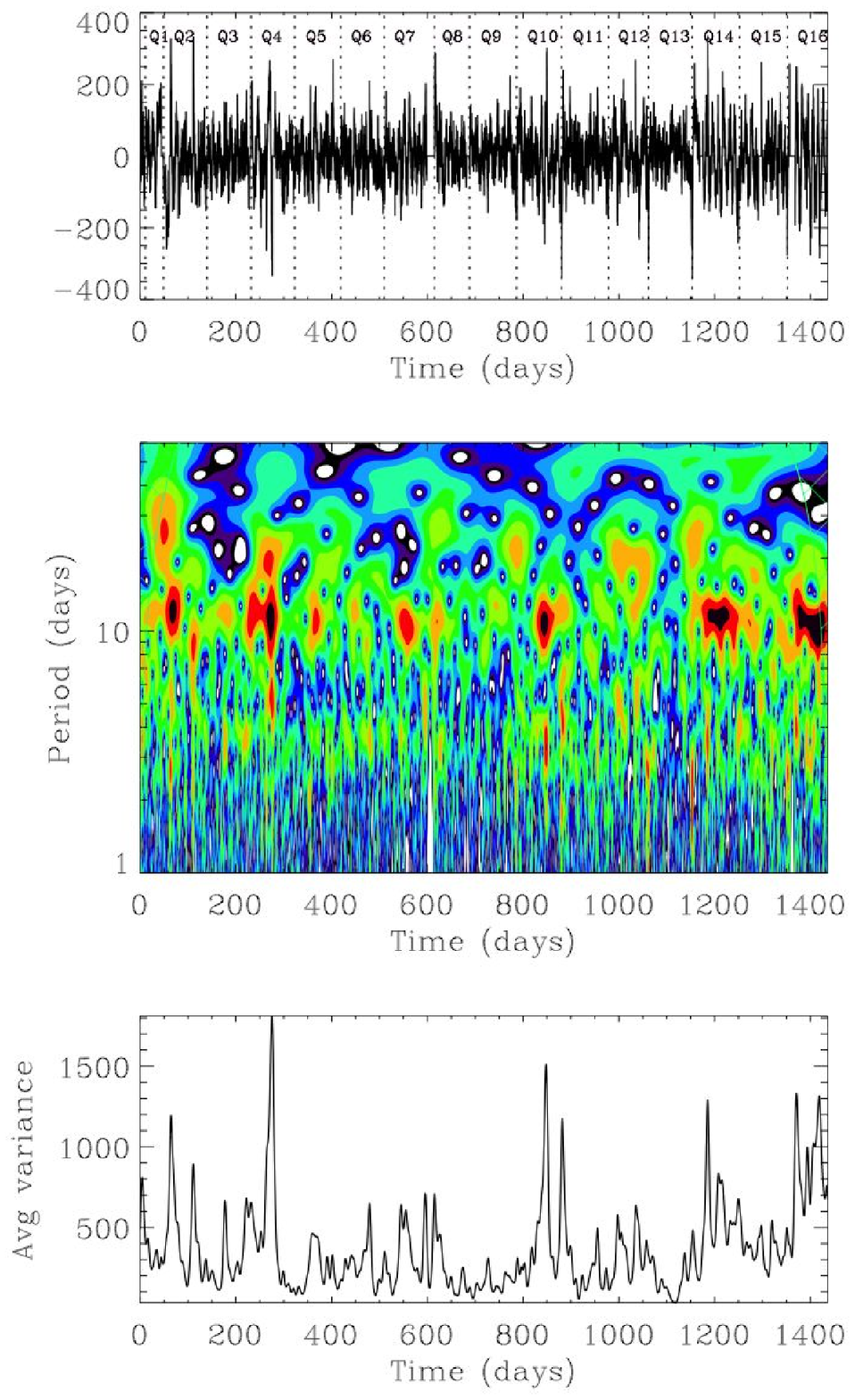}
\caption{ Wavelet analysis of KIC~7940546 {\bf (group O)}. Same representation as in Figure~\ref{wavelets1}. The magnetic proxy (bottom panel) was computed between periods of 3 and 18 days.}
\label{wavelets11}
\end{center}
\end{figure}

\begin{figure}[htbp]
\begin{center}
\includegraphics[width=10cm, trim=0.5cm 6cm 4cm 0.5cm]{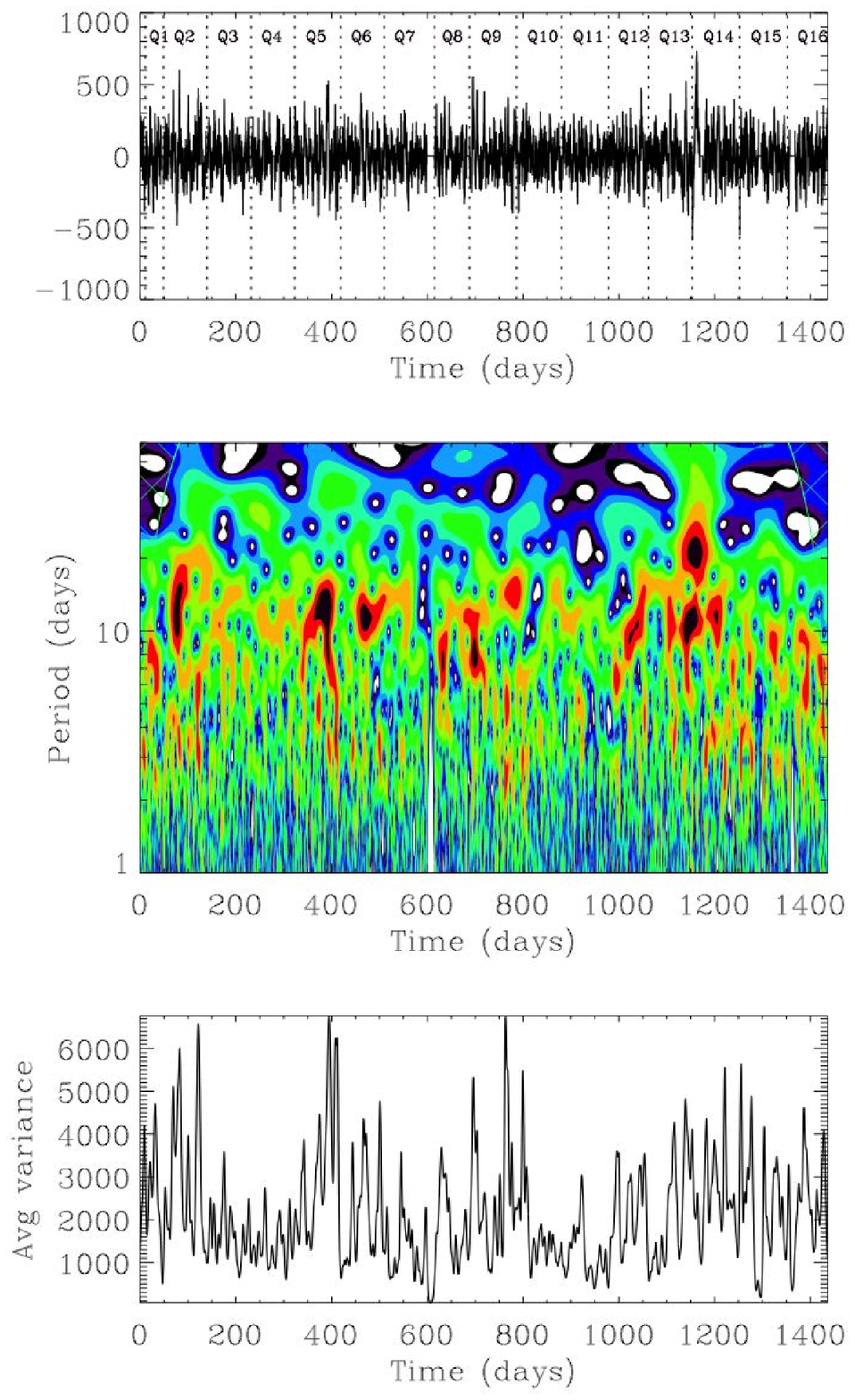}
\caption{ Wavelet analysis of KIC~8026226 {\bf (group O)}. Same representation as in Figure~\ref{wavelets1}. The magnetic proxy (bottom panel) was computed between periods of 2 and 16 days.}
\label{wavelets12}
\end{center}
\end{figure}

\begin{figure}[htbp]
\begin{center}
\includegraphics[width=10cm, trim=0.5cm 6cm 4cm 0.5cm]{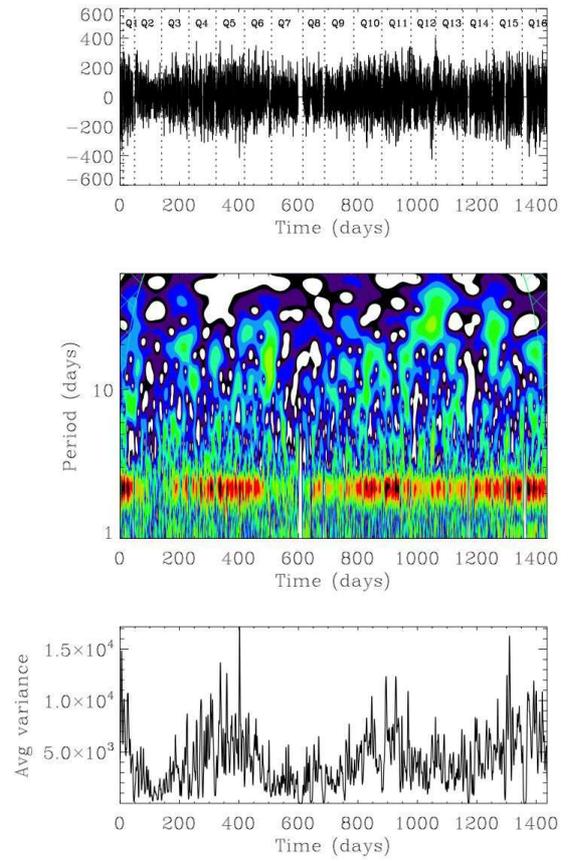}
\caption{ Wavelet analysis of KIC~9226926 {\bf (group L)}. Same representation as in Figure~\ref{wavelets1}. The magnetic proxy (bottom panel) was computed between a period of 1 and 4 days.}
\label{Kaa_wavelets_bis}
\end{center}
\end{figure}

\begin{figure}[htbp]
\begin{center}
\includegraphics[width=10cm, trim=0.5cm 6cm 4cm 0.5cm]{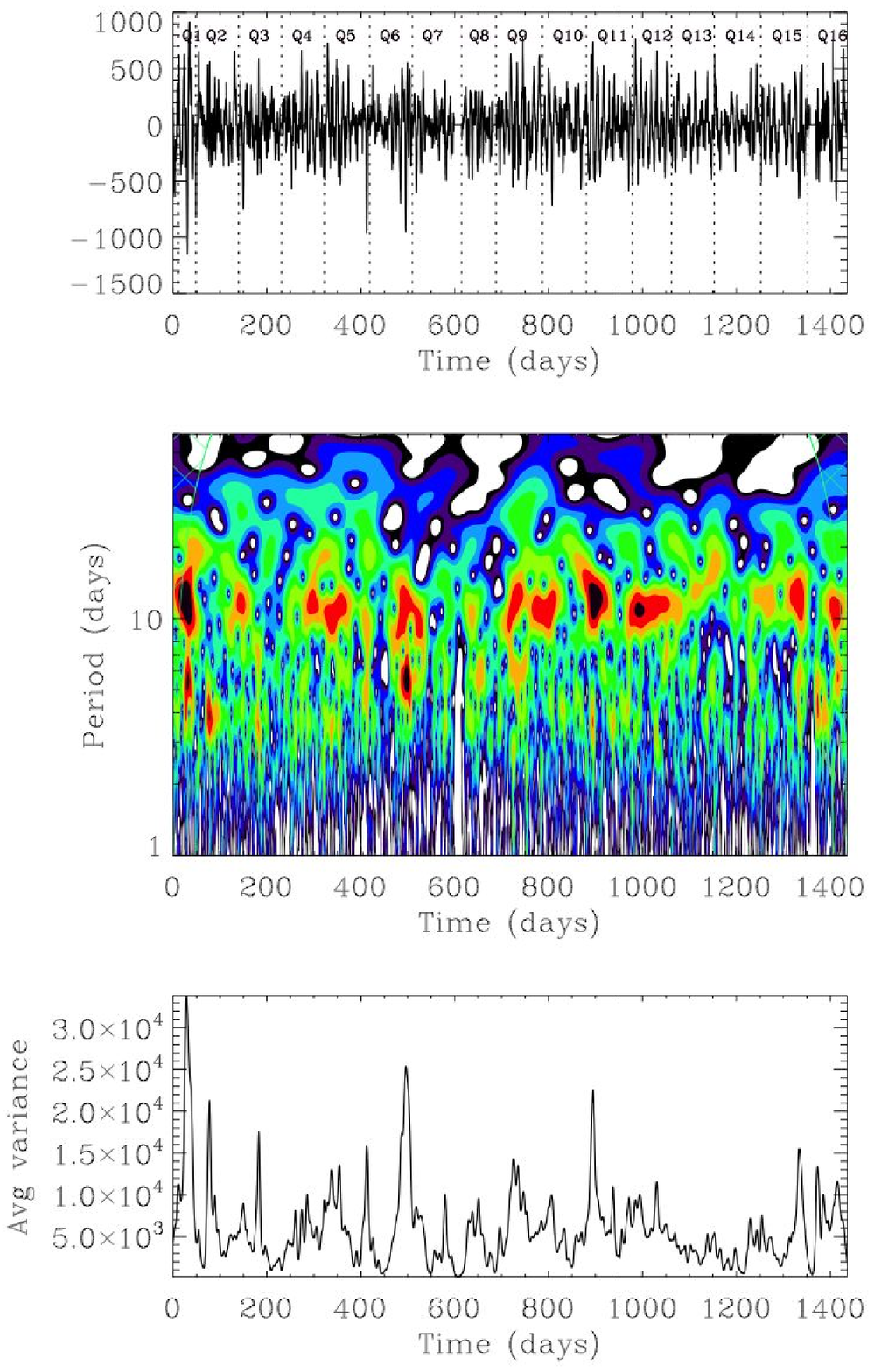}
\caption{ Wavelet analysis of KIC~9289275 {\bf (group O)}. Same representation as in Figure~\ref{wavelets1}. The magnetic proxy (bottom panel) was computed between periods of 2.3 and 17 days.}
\label{wavelets14}
\end{center}
\end{figure}

\begin{figure}[htbp]
\begin{center}
\includegraphics[width=10cm, trim=0.5cm 6cm 4cm 0.5cm]{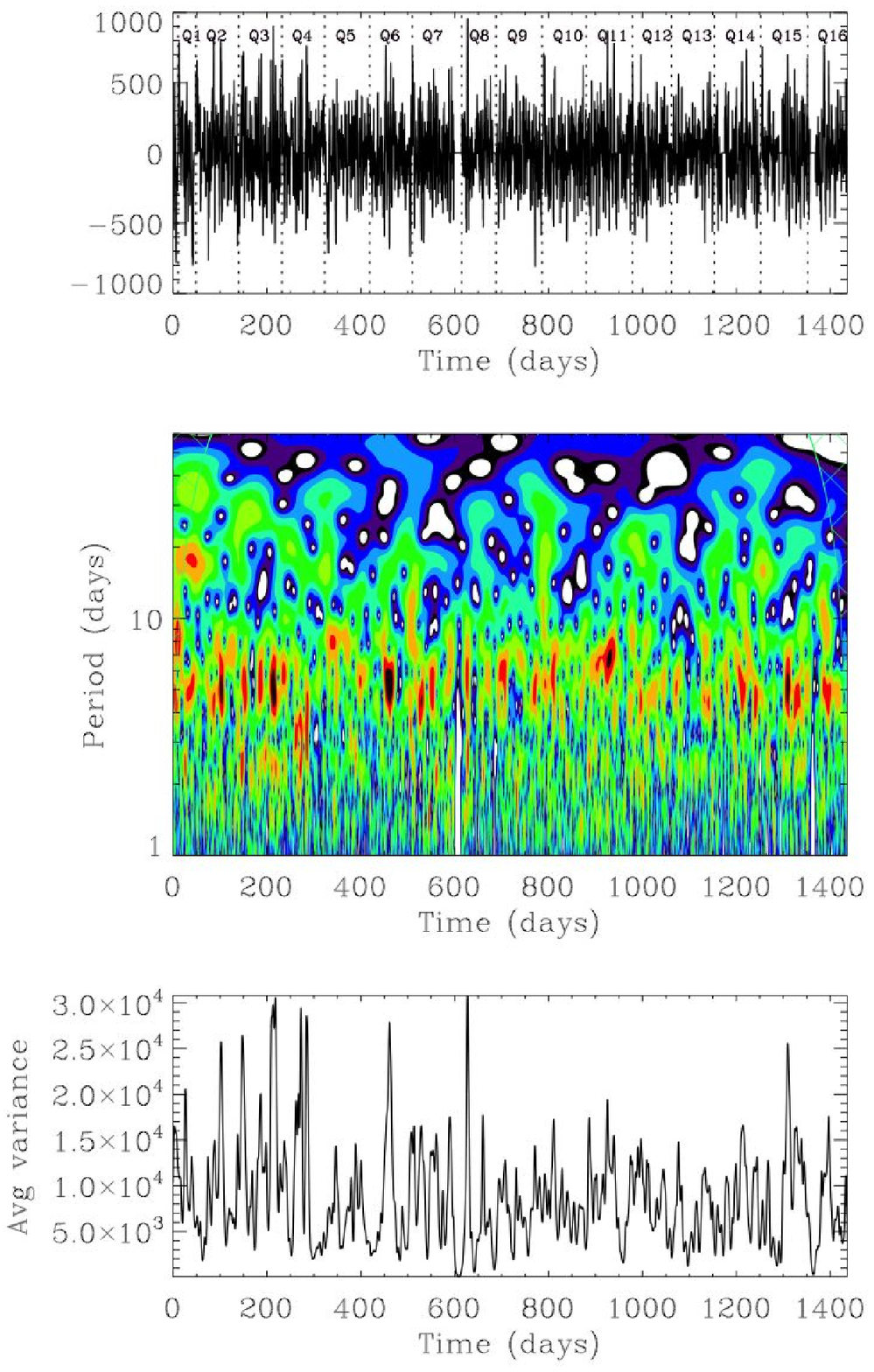}
\caption{ Wavelet analysis of KIC~9812850 {\bf (group O)}. Same representation as in Figure~\ref{wavelets1}. The magnetic proxy (bottom panel) was computed between periods of 1.8 and 10 days.}
\label{wavelets15}
\end{center}
\end{figure}

\begin{figure}[htbp]
\begin{center}
\includegraphics[width=10cm, trim=0.5cm 6cm 4cm 0.5cm]{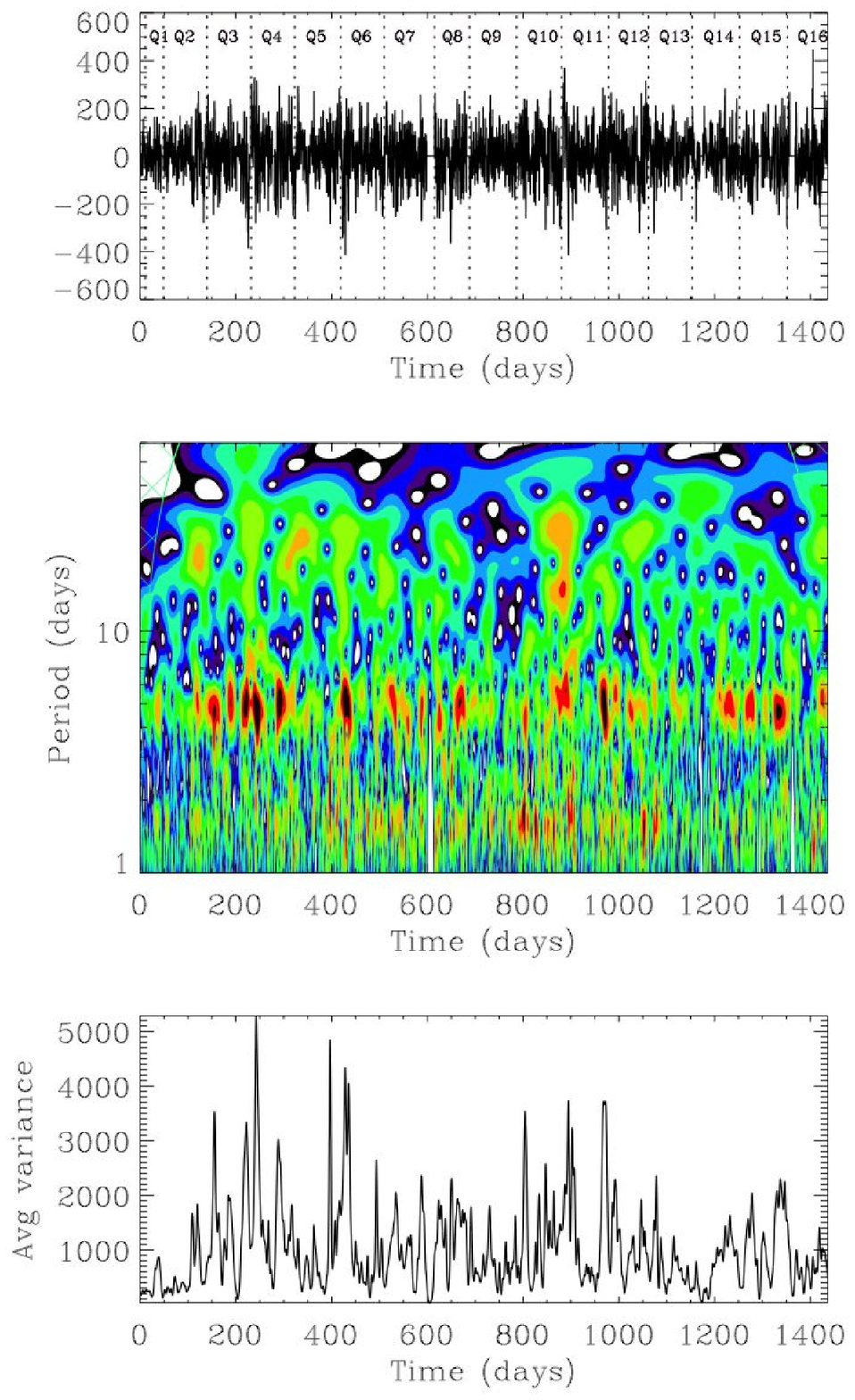}
\caption{ Wavelet analysis of KIC~10016239 {\bf (group O)}. Same representation as in Figure~\ref{wavelets1}. The magnetic proxy (bottom panel) was computed between periods of  1.8 and 10 days.}
\label{wavelets16}
\end{center}
\end{figure}

\begin{figure}[htbp]
\begin{center}
\includegraphics[width=10cm, trim=0.5cm 6cm 4cm 0.5cm]{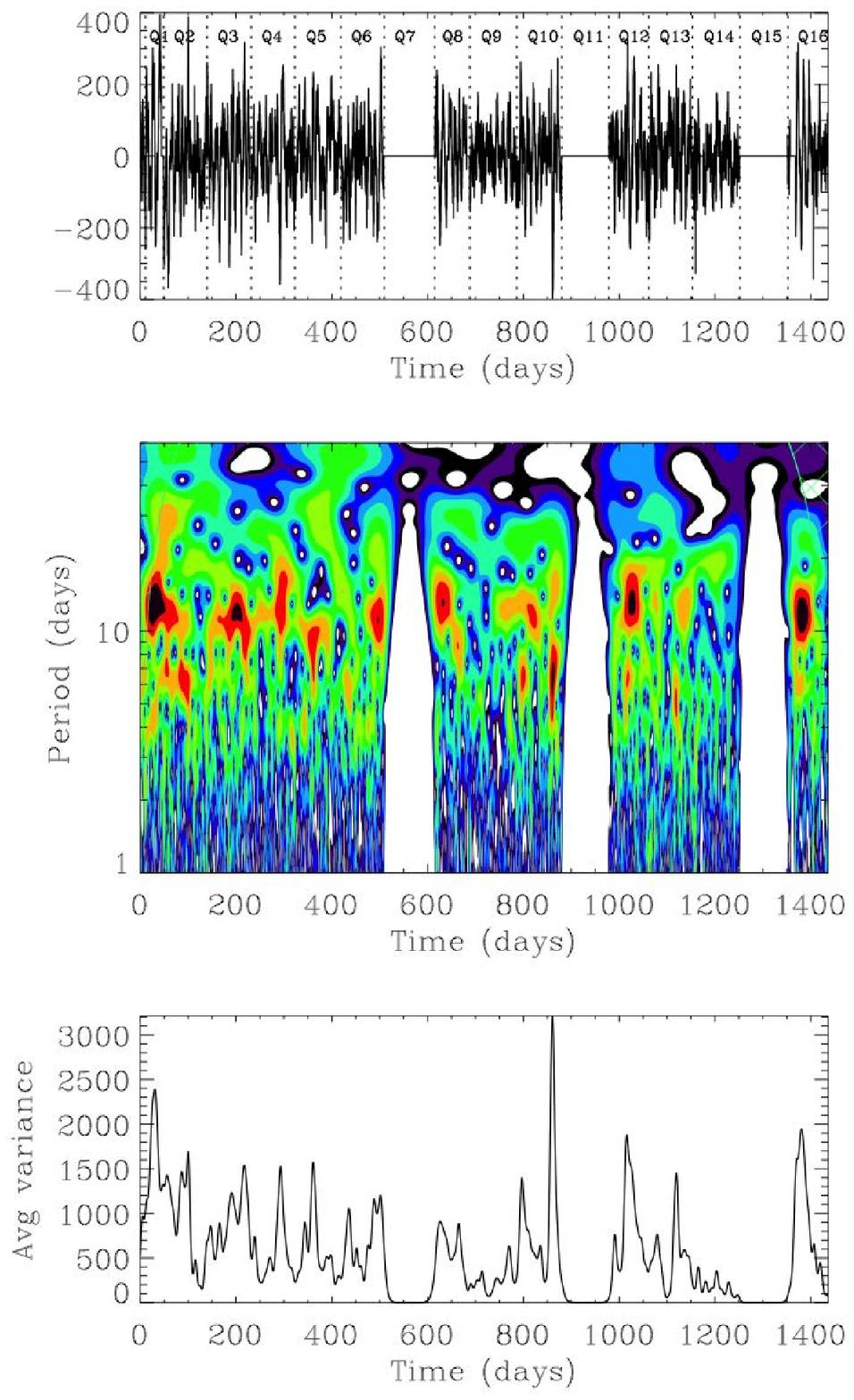}
\caption{ Wavelet analysis of KIC~10162436 {\bf (group O)}. Same representation as in Figure~\ref{wavelets1}. The magnetic proxy (bottom panel) was computed between periods of 3.8 and 17 days.}
\label{wavelets16}
\end{center}
\end{figure}
\clearpage

\begin{figure}[htbp]
\begin{center}
\includegraphics[width=10cm, trim=0.5cm 6cm 4cm 0.5cm]{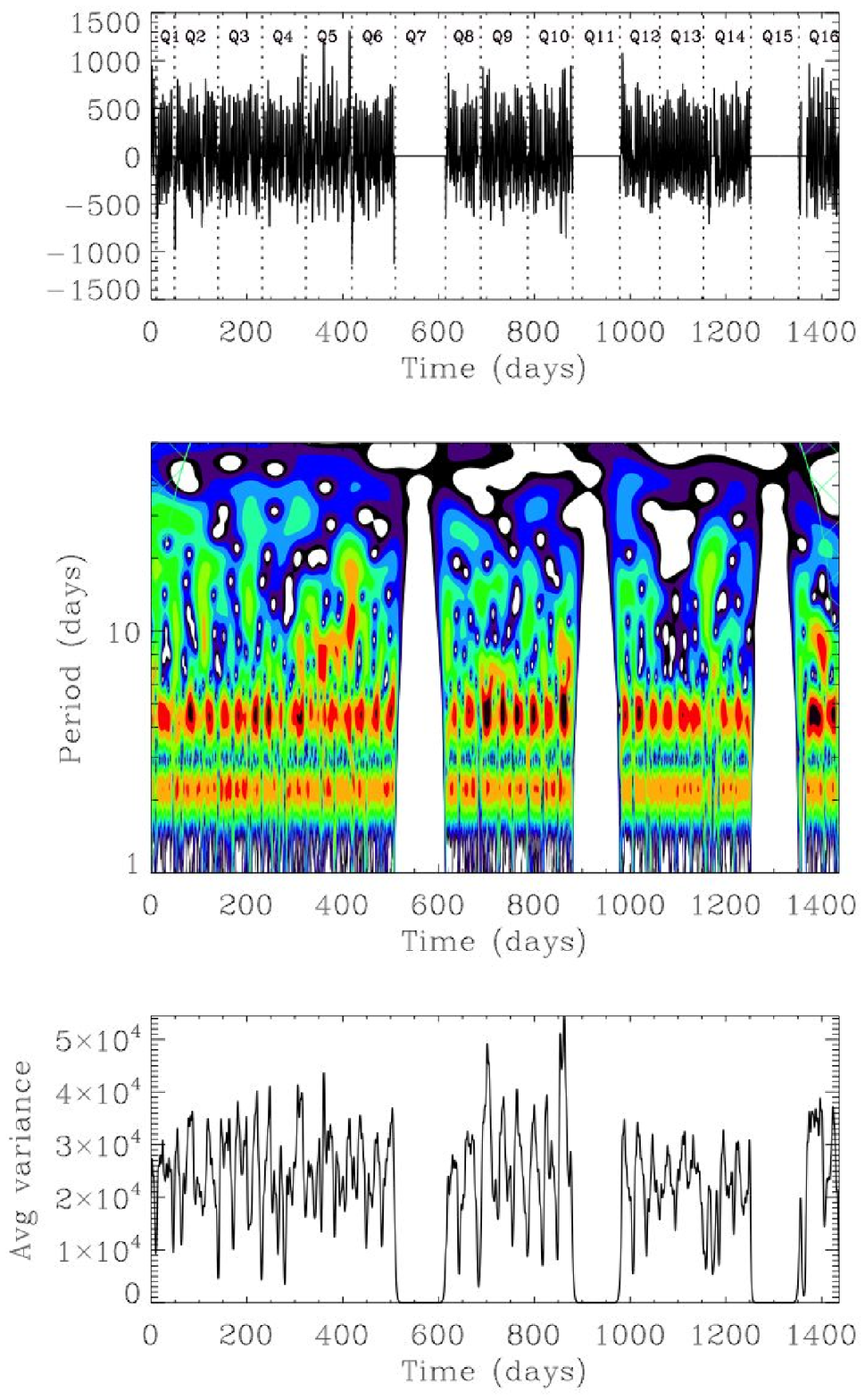}
\caption{ Wavelet analysis of KIC~10355856 {\bf (group L)}. Same representation as in Figure~\ref{wavelets1}. The magnetic proxy (bottom panel) was computed between periods of 1.7 and 8 days.}
\label{wavelets17}
\end{center}
\end{figure}

\begin{figure}[htbp]
\begin{center}
\includegraphics[width=10cm, trim=0.5cm 6cm 4cm 0.5cm]{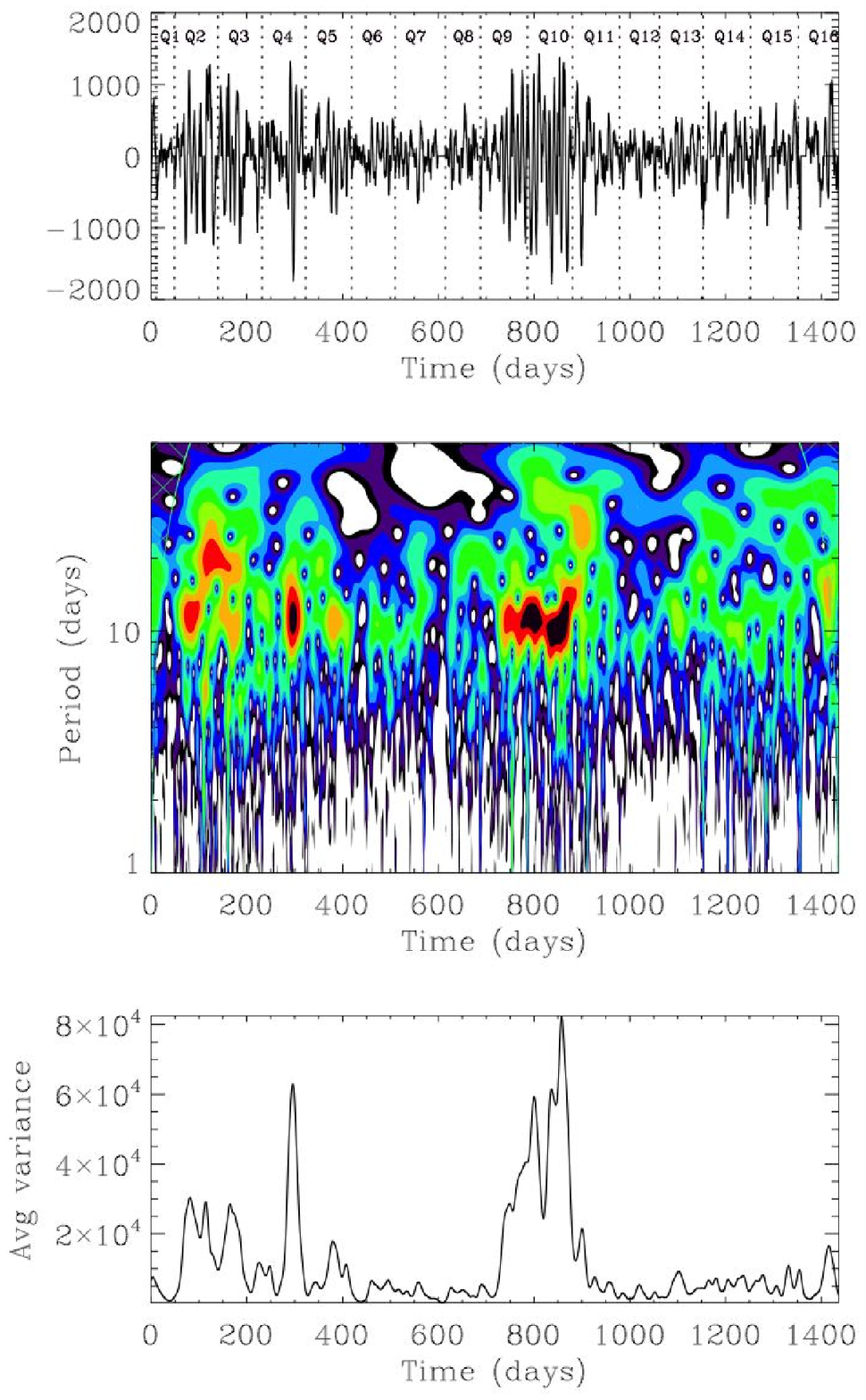}
\caption{ Wavelet analysis of KIC~10644253 {\bf (group C)}. Same representation as in Figure~\ref{wavelets1}. The magnetic proxy (bottom panel) was computed between periods of 5 and 18 days.}
\label{wavelets18}
\end{center}
\end{figure}

\begin{figure}[htbp]
\begin{center}
\includegraphics[width=10cm, trim=0.5cm 6cm 4cm 0.5cm]{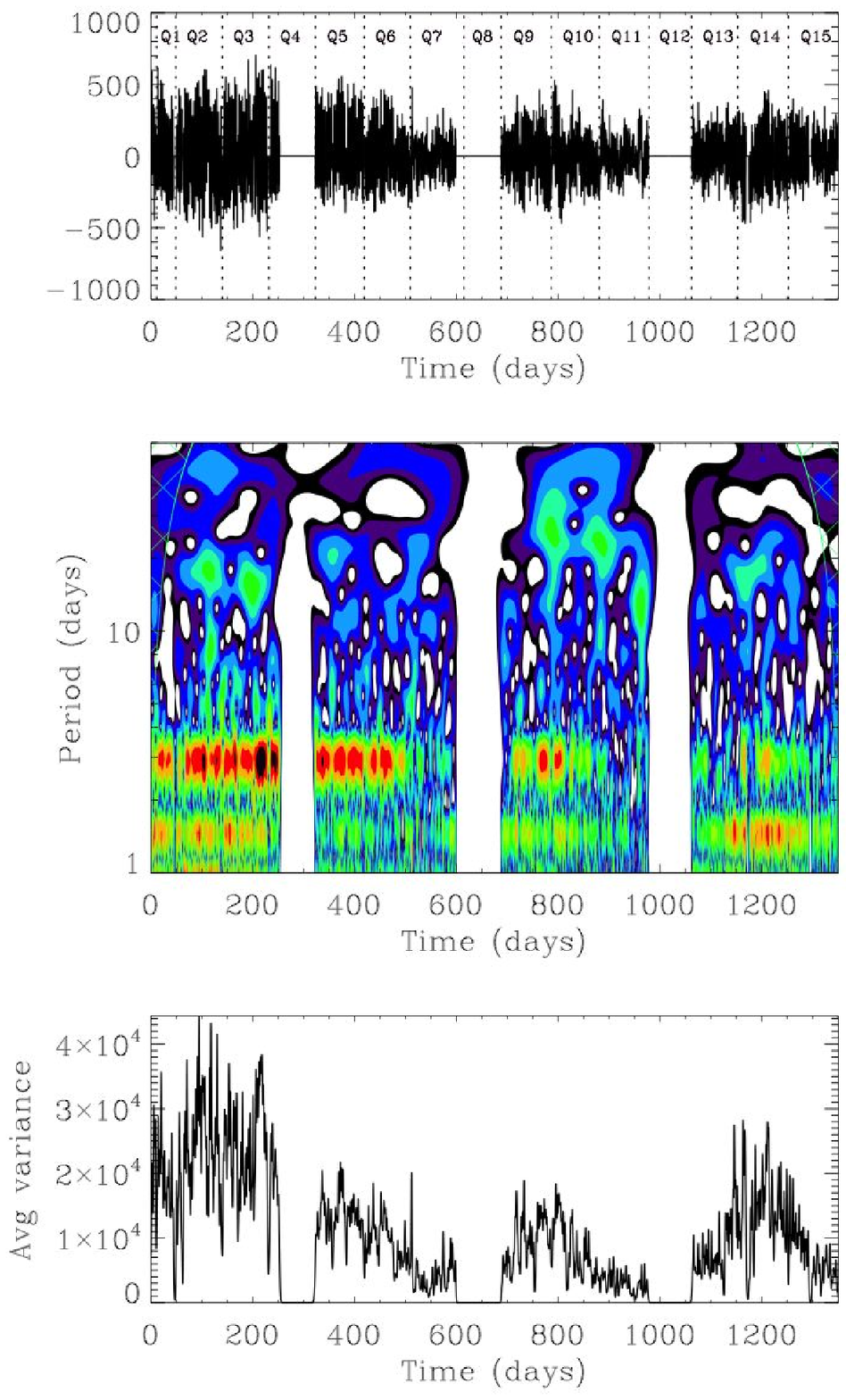}
\caption{ Wavelet analysis of KIC~11070918 {\bf (group L)}. Same representation as in Figure~\ref{wavelets1}. The magnetic proxy (bottom panel) was computed between periods of 0.8 an 5 days.}
\label{wavelets19}
\end{center}
\end{figure}

\begin{figure}[htbp]
\begin{center}
\includegraphics[width=10cm, trim=0.5cm 6cm 4cm 0.5cm]{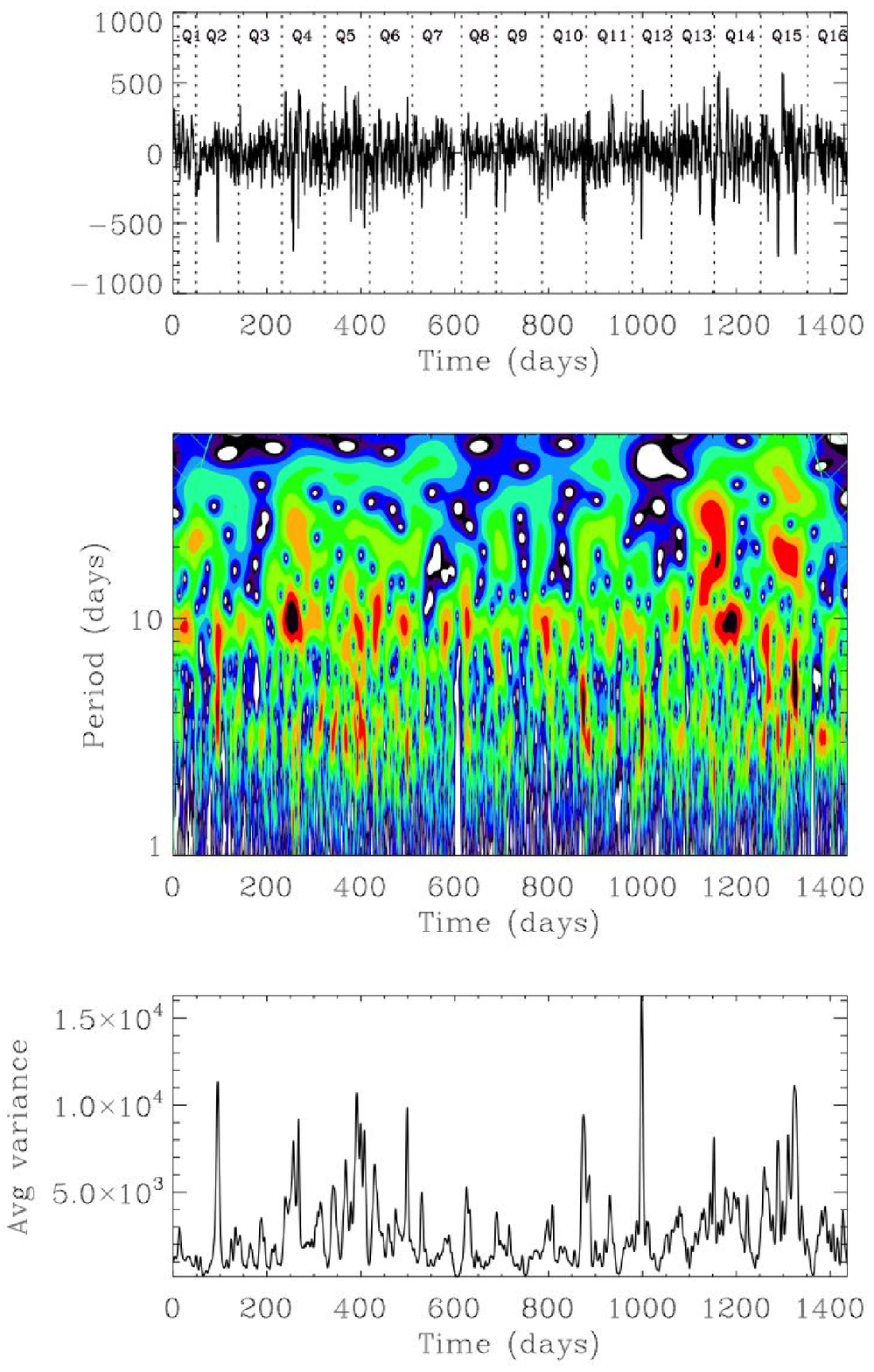}
\caption{ Wavelet analysis of KIC~12009504 {\bf (group T)}. Same representation as in Figure~\ref{wavelets1}. The magnetic proxy (bottom panel) was computed between periods of 1.8 and 15 days.}
\label{wavelets20}
\end{center}
\end{figure}

\end{document}